\documentclass[11pt]{article}
\usepackage{geometry}
\usepackage{amsmath,amsfonts}
\usepackage[dvipsnames]{xcolor}
\usepackage{empheq,tensor,cancel,braket,tcolorbox,blkarray,authblk,cite}
\usepackage[colorlinks=true, linkcolor=blue, citecolor=blue, linktoc=all]{hyperref}
\usepackage{bm}
\usepackage{tabularx}
\usepackage{ltablex}
\usepackage{subfigure}

\usepackage{tikz}
\usepackage{cleveref}
\usepackage{supertabular}
\usepackage{todonotes}
\usepackage{mathrsfs}
\usepackage{array}
\usepackage{lipsum}
\usepackage{breqn}
\usepackage{physics}
\usepackage{tikz-cd}
\usepackage{enumerate}
\usepackage{pifont}
\usepackage{float} 
\usepackage{helvet} 
\usepackage{amsthm} 
\usepackage{amssymb} 
\usepackage{booktabs} 
\usepackage{spectralsequences} 

\newcommand{\newreptheorem}[2]{%
\newenvironment{rep#1}[1]{%
 \def\rep@title{#2 \ref{##1}}%
 \begin{rep@theorem}}%
 {\end{rep@theorem}}}
\makeatother

\newreptheorem{lemma}{Lemma}

\newreptheorem{conj}{Conjecture}

\usepackage{lscape} 

\usepackage[T1]{fontenc} 

\DeclareMathOperator{\gh}{gh} 

\newcommand{\un}[1]{\underline{#1}}
\newcommand{\mX}{\mathfrak{X}}
\newcommand{\mY}{\mathfrak{Y}}

\newcommand{\mg}{\mathfrak{g}}
\newcommand{\umX}{\underline{\mathfrak{X}}}
\newcommand{\umY}{\underline{\mathfrak{Y}}}
\newcommand{\umZ}{\underline{\mathfrak{Z}}}
\newcommand{\uX}{\underline{X}}
\newcommand{\uY}{\underline{Y}}
\newcommand{\uZ}{\underline{Z}}
\newcommand{\uE}{\underline{E}}

\newcommand{\umu}{\underline{\mu}}
\newcommand{\unu}{\underline{\nu}}
\newcommand{\End}{\text{End}}

\newcommand{\nn}{\nonumber}

\newcommand{\p}{\partial}
\newcommand{\td}{{\rm d}}
\newcommand{\ts}{{\rm s}}
\newcommand{\lie}{\mathcal{L}}

\newcommand{\be}{\begin{equation}}
\newcommand{\ee}{\end{equation}}
\newcommand{\ba}{\begin{aligned}}
\newcommand{\ea}{\end{aligned}}
\newcommand{\bea}{\begin{eqnarray}}
\newcommand{\eea}{\end{eqnarray}}

\def\bp{\begin{pmatrix}}
\def\ep{\end{pmatrix}}

\def\im{\mathrm{im}}
\def\ker{\mathrm{ker}}

\allowdisplaybreaks[1]

\title{\huge On Quantum Aspects of 1-Form Symmetries I: BV--BRST Cohomology and Anomaly Polynomials}

\author[1]{Weizhen Jia\thanks{weizhenjia@cuhk.edu.hk}}
\author[2,3]{Yi-Nan Wang\thanks{ynwang@pku.edu.cn}}
\author[4]{Yi Zhang\thanks{yi.zhang@ipmu.jp}}

\affil[1]{Department of Physics, The Chinese University of Hong Kong, Sha Tin, Hong Kong, China}
\affil[2]{School of Physics, Peking University, Beijing, China, 100871}
\affil[3]{Center for High Energy Physics, Peking University, Beijing, China, 100871}
\affil[4]{Kavli IPMU (WPI), UTIAS, The University of Tokyo, Kashiwa, Chiba 277-8583, Japan}

\date{}							

\begin{document}
\maketitle

\begin{abstract}
We investigate the quantum aspects of gauging continuous 1-form global symmetries. In this paper, we study the BV--BRST quantization of a $U(1)$ 2-form gauge field, described geometrically by a $U(1)$ gerbe. Starting from the local \v Cech data of the gerbe, we construct the corresponding infinitesimal symmetry structure in terms of a Lie 2-algebroid, and show that, together with the associated exact Courant algebroid, it provides a natural geometric framework for the BV--BRST complex of this higher-form gauge theory. In this formulation, the field-ghost tower is encoded directly in the local gerbe data, and the higher Russian formula arises naturally from the relations among the connective structure, the curving, and the 3-form curvature. We further show that the resulting \v Cech--de~Rham bicomplex provides a natural setting for anomaly descent for $U(1)$ 1-form symmetries, and illustrate the construction with explicit examples in Maxwell theory.

\end{abstract}
~~\\
\begin{center}
\it --- Dedicated to the memory of Professor Rob Leigh ---
\end{center}

\newpage
\begingroup
\hypersetup{linkcolor=black}
\tableofcontents
\endgroup

\newpage

\section{Introduction}

Higher-form symmetries \cite{Gaiotto:2014kfa,Gomes:2023ahz,Bhardwaj:2023kri,Luo:2023ive,Iqbal:2024pee,Kaidi:2026urc} have become a standard part of the modern language of quantum field theory. In general, $p$-form global symmetries act on extended operators of dimension $p$ and are probed by background gauge fields of degree $p+1$. For $p \geqslant 1$, higher-form symmetries are necessarily Abelian, since they are realized by topological defects that can slide past one another without affecting correlation functions. The simplest example is a 1-form symmetry, which acts on line operators and couples to a background 2-form gauge field. Unlike the gauge fields associated with ordinary (0-form) symmetries---geometrically described by connections on principal bundles---the background gauge field for a 1-form symmetry is naturally formulated in terms of a \emph{gerbe}.\footnote{Non-Abelian higher gauge theory and various notions of non-Abelian gerbes have also been studied in the literature. We will not consider the non-Abelian case here.} This gerbe viewpoint is also the traditional entry point to higher gauge theory and to the geometry of $B$-fields in string theory \cite{Freed:1999vc,Kapustin:1999di,Figueroa-OFarrill:2000lcd,Bouwknegt:2001vu,Aschieri:2004yz,BaezSchreiber2005,Baez:2010ya}.

The gauging of higher-form symmetries is usually discussed at the classical action level, for example the gauging of electric and magnetic $U(1)$ 1-form symmetries in $4d$ Maxwell theory~\cite{Gaiotto:2014kfa}. In this case, the mixed 't Hooft anomaly between the electric and magnetic $U(1)$ 1-form symmetries can be viewed as a classical effect. Similar perspectives were taken to formulate charged matter under $U(1)$ higher-form symmetries in the mean string field theory and its generalizations, to build up Landau--Ginzburg models for higher symmetries as classical, effective models in loop and brane spaces~\cite{Iqbal_2022,Hidaka:2023gwh,Liu:2024znj,Kawana:2025vbi,Kawana:2026lwu}. A natural direction to investigate is the quantum aspects of the 1-form symmetry, i.e., are there new structures in the quantization of 1-form symmetries, and does it have quantum anomalies beyond the classical framework of anomaly polynomials. These are the main goals of this paper and the companion paper \cite{PartII}.

From the viewpoint of quantization, gauge theories with 2-form gauge fields $B$ must be treated using the Batalin--Vilkovisky (BV) formalism \cite{Batalin:1981jr,Batalin:1984jr,Henneaux:1992ig,Gomis:1994he}. The reason is that 2-form gauge theories possess gauge-for-gauge transformations, indicating that the gauge generators are not all independent. Such theories are therefore called \emph{reducible}, and the standard Faddeev--Popov procedure is not applicable. In the BV formalism, the space of fields is extended by introducing a 1-form ghost $C$ and a 0-form ghost-for-ghost $c$, together with their antifields. These fields are organized into a graded complex equipped with an odd, nilpotent differential, forming the BV--BRST complex. This complex also provides the natural cohomological framework for the descent equations and for the analysis of perturbative anomalies. 

For ordinary gauge theory, the corresponding BRST construction is encoded in the \emph{Russian formula}~\cite{Stora:1983ct}. If $A$ is a connection 1-form and $c$ is the ghost field, one combines the de~Rham differential $\td$ with the BRST operator $\ts$ and introduces the extended connection $A+c$. The Russian formula states that the corresponding extended curvature coincides with the ordinary curvature $F=\td A+\frac{1}{2}[A,A]$, namely
\begin{equation}
(\td+\ts)(A+c)+\frac{1}{2}[A+c,A+c]=F\,.
\end{equation}
Expanding this identity reproduces the BRST transformation laws of $A$ and $c$. It is also the basic starting point for the descent equations and hence for the cohomological treatment of anomalies \cite{wess1971consequences,Stora1977,Stora:1983ct,Zumino:1983rz,Zumino:1984ws,zumino1985chiral}.

This structure for ordinary gauge symmetries also admits an intrinsic geometric interpretation. In previous work \cite{Jia:2023tki,Ciambelli:2021ujl}, the BRST complex was geometrized using the \emph{Atiyah Lie algebroid} associated with a principal $G$-bundle \cite{Atiyah:1957}. In a local trivialization, the algebroid coboundary operator reproduces the de~Rham differential together with the Chevalley--Eilenberg differential, the latter being identified with the BRST operator and its ghost-number grading. In this language, the Russian formula emerges naturally from the horizontality of the algebroid curvature. Moreover, characteristic classes on the algebroid produce the descent equations, and the consistent anomaly is identified cohomologically as the ghost number one sector of the BRST complex. In this sense, neither the BRST complex nor the Russian formula is an external input: both are encoded in the intrinsic geometry of the Atiyah sequence and become manifest upon trivialization.

One of the main purposes of the present paper is to extend this algebroid construction from principal bundles to gerbes, and thereby to place the BV--BRST complex of a 2-form gauge field on the same conceptual footing. A key point is that, unlike the Atiyah Lie algebroid, the higher structure associated with a gerbe does not naturally arise from an underlying vector bundle. For this reason, the local patching data of the gerbe provide the natural framework. In particular, we show that the BRST complex can be recast as a \v Cech--de~Rham bicomplex, in which one differential is the de~Rham differential $\td$ and the other records the \v Cech, or equivalently the ghost number grading. This formulation is especially convenient for carrying over to the gerbe case, where the geometry is naturally described in terms of local data and their gluing relations.

Infinitesimal symmetries of gerbes have been studied extensively in the literature. More broadly, higher gauge theory admits several complementary formalisms, such as $L_\infty$-algebras, principal 2-bundles (and higher bundles), string bundles, and related higher-categorical structures~\cite{Baez0307,Baez:2003fs,BaezSchreiber2005,sati2009algebra,Baez:2010ya,Fiorenza:2012tb,Jurco:2018sby,Kristel:2022uty,Kristel:2022maw,Chen:2022hct,Kristel:2023gus,Bunk:2023sov,Rist:2023zsl}. In this paper we follow the higher algebroid approach by Collier \cite{collier2011}, which formulates the infinitesimal symmetry structure of a gerbe in sheaf language as a higher analogue of the Atiyah Lie algebroid.\footnote{Various other notions of higher Lie algebroid as generalizations of the Atiyah Lie algebroid have also been proposed in the math literature, see, e.g., \cite{Sheng_2017,Sheng_2018,Chatterjee_2022,Jotz_Lean_2019}.} We work with the corresponding \v Cech presentation of this Lie 2-algebroid, which keeps the local patching and gluing data explicit. In this presentation, the physicists' familiar BV fields $(B,C,c)$ and their ghost numbers appear transparently, in close parallel with the trivialized Lie algebroid description of the BRST complex for ordinary gauge symmetry.

Nevertheless, the Lie 2-algebroid by itself is not yet the whole story: the higher Russian formula involves not only the connective structure but also the curving and the resulting globally defined curvature 3-form $H$. Geometrically, a gerbe with connective structure canonically determines an \emph{exact Courant algebroid}~\cite{courant1988beyond,courant1990dirac,liu1997manin,roytenberg1999courant,Bressler:2002ur}, and a choice of curving provides a natural (isotropic) split of the Courant sequence. Courant algebroids are the central structure of generalized geometry \cite{Hitchin:2003cxu,Gualtieri:2003dx,Hitchin:2005in,Hitchin:2010qz,gualtieri2011generalized}, which is a framework that has appeared in physics in contexts such as T-duality, double field theory, and the description of NS-NS $B$-fields in type~II string theory~\cite{Grana:2004bg,Grana:2005sn,Grana:2008yw,Hull:2009mi,Gabella:2010laf,Hohm:2010pp,Berman:2010is,Hohm:2010xe,Hohm:2013bwa,Geissbuhler:2013uka}. We show that it is the combination of the Lie 2-algebroid of infinitesimal symmetries together with the associated exact Courant algebroid that yields the higher Russian formula
\begin{align}
(\td+\ts)(B-C+c)=H\,,
\end{align} 
and the resulting \v Cech--de~Rham bicomplex naturally serves as BV--BRST complex. In particular, the tower $(B,C,c)$ and the higher Russian formula are consequences of the Lie 2-algebroid/Courant-algebroid package derived from a gerbe with connective structure, rather than a constraint imposed to engineer the BV--BRST differential.\footnote{Courant algebroids are also often viewed as a higher analogue of the Atiyah Lie algebroid for a gerbe. However, the Courant algebroid by itself does not encode the full BV tower $(B,C,c)$; for our purposes the Lie 2-algebroid captures the infinitesimal symmetries of the gerbe, while the associated Courant algebroid supplies the additional geometric input needed to produce the higher Russian formula.} 

Altogether, this leads to the following correspondence between the physical description of gauge symmetry and its geometric formulation in Table~\ref{table1}:
\begin{table}[H]
\centering
\begin{tabular}{p{8cm}|p{5cm}}
\toprule
Physical description & Geometric formulation \\
\midrule
Classical gauge theory with 0-form symmetry & Principal bundle \\
Quantum gauge theory with 0-form symmetry & Atiyah Lie algebroid \\
Classical gauge theory with 1-form symmetry & Gerbe \\
Quantum gauge theory with 1-form symmetry & Lie 2-algebroid and associated Courant algebroid \\
\bottomrule
\end{tabular}
\caption{Physics/geometry correspondence for ordinary and higher-form gauge symmetry.}
\label{table1}
\end{table}

Having obtained this BV--BRST bicomplex geometrically, we apply it to study the anomalies of 1-form symmetries. As in the 0-form symmetry case, characteristic classes built from the curvature data define Chern--Simons forms whose expansion in form degree and ghost number produces descent equations and the Wess--Zumino consistency condition. The characteristic classes play the role of anomaly polynomials, and the consistent anomaly is identified as the ghost number one component in the descent.

This paper is organized as follows. In Section~\ref{sec:Atiyah}, we review the Atiyah Lie algebroid of a principal $U(1)$-bundle and explain how, after local trivialization and in its \v Cech presentation, it reproduces the BRST complex and the ordinary Russian formula for 0-form gauge symmetry. In Section~\ref{sec:Lie2Alg}, we pass to the higher setting and construct the \v Cech presentation of the Lie $2$-algebroid associated to a $U(1)$-gerbe, showing that its split is encoded by the connective structure and leads to a higher analogue of the Russian formula. In Section~\ref{sec:Courant}, we introduce the exact Courant algebroid canonically associated to a gerbe with connective structure and show that its isotropic split is determined by the curving. In Section~\ref{sec:Russian}, we combine these two geometric structures to derive the full higher Russian formula and interpret the resulting \v Cech--de~Rham bicomplex physically as the BV--BRST complex of the gerbe gauge theory. In Section~\ref{sec:descent-higherform}, we apply this bicomplex to anomaly descent for 1-form symmetries, formulate the corresponding anomaly polynomials, and discuss examples in $4d$ and $5d$ Maxwell theories. We conclude in Section~\ref{sec:conclusions}. In Appendix~\ref{App:ALACourant}, we provide background on Atiyah and Courant algebroids, while Appendix~\ref{App:BV} reviews the BV formalism for Abelian 2-form gauge fields.

\section{Atiyah Lie Algebroids and the BRST Complex}
\label{sec:Atiyah}
In this section we review the Atiyah Lie algebroid associated to a principal bundle. We will focus on  the fact that, after local trivialization, its exterior algebra reproduces the BRST complex and the Russian formula for ordinary (0-form) gauge symmetries. We then reformulate the same structure in \v Cech language. The latter description will be particularly useful for our purposes, since it generalizes directly to the gerbe case and leads naturally to the higher Russian formula for the BV formalism. For more basics of Atiyah Lie algebroids, see Appendix~\ref{App:ALA}. A comprehensive introduction to the theory of Lie groupoids and Lie algebroids is \cite{mackenzie2005general}. For detailed introduction on the role of Atiyah Lie algebroid in the BRST formalism, the reader may refer to \cite{Jia:2023tki,Ciambelli:2021ujl} and thesis \cite{Jia:2024ujz}.

\subsection{Atiyah Lie Algebroids and Their Trivializations}
\label{sec:ALAtriv}

Let $P(M,G)$ be a principal $G$-bundle over the base manifold $M$, with Lie algebra $\mg=\mathrm{Lie}(G)$. The \emph{Atiyah Lie algebroid} derived from $P$ is the transitive Lie algebroid represented by the following short exact sequence:
\begin{equation} \label{ALA}
\begin{tikzcd}
0
\arrow{r} 
& 
L = P\times_{\mathrm{Ad}} \mathfrak{g}
\arrow{r}{j} 
&
A = TP/G
\arrow{r}{\rho} 
& 
TM
\arrow{r} 
&
0\,.
\end{tikzcd}
\end{equation}
Here $L$ is the adjoint bundle of Lie algebras, called the isotropy bundle. The image of $L$ under $j$ defines the vertical subbundle of $A$, while the anchor map $\rho$ projects onto the tangent bundle of the base. In physical terms, sections of $L$ generate local gauge transformations, whereas sections of $TM$ generate local diffeomorphisms. Thus, the Atiyah Lie algebroid characterizes the infinitesimal symmetries of the principal bundle $P$, with the internal and spacetime symmetries implemented on an equal footing.

\paragraph{Connection and split.}
A \emph{connection} on the Atiyah Lie algebroid is specified by a choice of \emph{split} $\sigma:TM\to A$ of the Atiyah sequence \eqref{ALA}, or equivalently by a horizontal subbundle complementary to the vertical one. It may also be encoded by a map
$\omega:A\to L$ called the \emph{connection reform}. Thus, we have a pair of maps
\begin{align}
\omega:A\to L\,,\qquad \sigma:TM\to A\,,
\end{align}
such that\footnote{The sign convention follows \cite{Ciambelli:2021ujl,Jia:2023tki,Jia:2024ujz,Fournel:2012uv,Lazzarini_2012}, so that the curvature of $-\omega$ in \eqref{Romega} is consistent with the curvature 2-form defined in \eqref{curvaturereform}.}
\begin{align}
\ker(\omega)=\im(\sigma)\,,\qquad \rho\circ\sigma=\mathrm{id}_{TM}\,,\qquad
\omega\circ j=-\mathrm{id}_L\,.
\end{align}
In this way, the exact sequence \eqref{ALA} also goes in the opposite direction:
\begin{equation}
\label{ALAsplit}
\begin{tikzcd}
0
\arrow{r} 
& 
L
\arrow{r}{j} 
\arrow[bend left]{l} 
& 
A
\arrow{r}{\rho} 
\arrow[bend left]{l}{\omega}
& 
TM
\arrow{r} 
\arrow[bend left]{l}{\sigma}
&
0\,.
\arrow[bend left]{l} 
\end{tikzcd}
\end{equation}
The image of $\sigma$ defines the horizontal subbundle $H\subset A$, and hence
\begin{align}
A=H\oplus V\,,
\end{align}
where $V=\im(j)$ is the vertical subbundle.

\paragraph{Local trivialization.}
Let $\{U_i\}$ be an open cover of $M$. Over each $U_i$, the Atiyah Lie algebroid admits a local trivialization
\begin{align}
\label{localtrivializedA}
\tau_i:A|_{U_i}\longrightarrow TU_i\oplus L_i\,,
\qquad
L_i\equiv L|_{U_i}\,,
\end{align}
which defines a \emph{trivialized algebroid} $A_\tau$. Strictly speaking, the trivialized algebroid is not simply the local direct sum $TU_i\oplus L_i$ on a single patch, but rather an algebroid obtained from $A$ by a local isomorphism
\begin{align}
\tau:A\longrightarrow A_\tau\,,
\end{align}
where $A_\tau$ is constructed by sewing together the local models $TU_i\oplus L_i$ on the open cover \cite{Jia:2023tki}. In this sense, the trivialized algebroid should be understood as a local representative of the isomorphism class of $A$. Since our discussion below is local, we will often suppress the distinction between $A_\tau|_{U_i}$ and $TU_i\oplus L_i$, but it should be kept in mind that the trivialization is globally defined and implemented by the algebroid isomorphism $\tau$.

Let $\{\uE_{\un\alpha},\uE_{\un A}\}$ be a local split basis for $A$, where $\{\uE_{\un\alpha}\}$ spans the horizontal subbundle and $\{\uE_{\un A}\}$ spans the vertical subbundle. In terms of a basis $\{\un\p_\mu\}$ of $TU_i$ and $\{\un t_A\}$ of $L_i$, these basis sections may be written as
\begin{align}
\uE_{\un\alpha}
=
\rho^\mu{}_{\un\alpha}(\un\p_\mu+b^A_{i,\mu}\un t_A)\,,
\qquad
\uE_{\un A}
=
\omega^A{}_{\un A}\un t_A\,,
\end{align}
where the local $\mg$-valued 1-form
\begin{align}
b_i=b^A_{i,\mu}\td x^\mu\otimes \un t_A\in\Omega^1(U_i,\mg)
\end{align}
encodes the horizontal lift. Physically, $b_i$ is the local gauge field associated to the principal connection.

A general section $\umX\in\Gamma(A|_{U_i})$ then decomposes as
\begin{align}
\umX
&=
\mX^{\un\alpha}\uE_{\un\alpha}+\mX^{\un A}\uE_{\un A}\nn\\
&=
\mX^{\un\alpha}\rho^\mu{}_{\un\alpha}(\un\p_\mu+b^A_{i,\mu}\un t_A)
+\mX^{\un A}\omega^A{}_{\un A}\un t_A\nn\\
&=
X^\mu\un\p_\mu+\mu_i^A\un t_A\,,
\end{align}
where
\begin{align}
X^\mu=\mX^{\un\alpha}\rho^\mu{}_{\un\alpha}\,,\qquad
\mu_i^A=X^\mu b^A_{i,\mu}+\tilde\mu_i^A\,, \qquad \tilde\mu_i^A=\mX^{\un A}\omega^A{}_{\un A}\,.
\end{align}
Thus, in a local trivialization over $U_i$, a section $\mX$ of the Atiyah Lie algebroid is represented by a pair
\begin{align}
(\uX,\umu_i)\in \Gamma(TU_i)\oplus\Gamma(L_i)\,,
\end{align}
where $\umu_i$ contains a horizontal piece $i_Xb_i$ and a vertical piece $\tilde\umu_i$.  In the trivialization, one also has the following Lie bracket:\footnote{In fact, the Atiyah Lie algebroid can be defined locally by the Lie bracket \eqref{ALiebracket} together with the patching condition \eqref{ALAintersect}.}
\begin{align} 
\label{ALiebracket}
[(\uX_i,\umu_i),(\uY_i,\unu_i)]_{TU_i\oplus L_i}
= [\uX_i,\uY_i]_{TU_i}+[\umu_i,\unu_i]_{L_i}+\lie_{X_i} \unu_i-\lie_{Y_i}\umu_i\,.
\end{align}

\paragraph{The split and the local gauge field.}
The local 1-form $b_i$ appearing above is not merely a convenient parametrization of the split basis: it is precisely the local gauge field of the principal connection. To see this, let us compare the local decomposition of a section with its patching law on overlaps.

On an overlap $U_{ij}$, a section of the Atiyah Lie algebroid transforms according to
\begin{align}
\label{ALAintersect}
(\uX,\umu_i)\sim (\uX,g_{ij}^{-1}\umu_jg_{ij}+i_Xg_{ij}^{-1}\td g_{ij})\,.
\end{align}
Substituting the decomposition
\begin{align}
\umu_i=i_X b_i+\tilde\mu_i
\end{align}
into \eqref{ALAintersect}, we obtain
\begin{align}
i_X b_i+\tilde\mu_i
=
g_{ij}^{-1}(i_X b_j+\tilde\mu_j)g_{ij}+i_Xg_{ij}^{-1}\td g_{ij}\,.
\end{align}
Since the horizontal and vertical pieces transform independently, it follows that
\begin{align}
\label{bmutransform}
b_i=g_{ij}^{-1}b_jg_{ij}+g_{ij}^{-1}\td g_{ij}\,,
\qquad
\tilde\mu_i=g_{ij}^{-1}\tilde\mu_jg_{ij}\,.
\end{align}
Thus, $b_i$ transforms exactly as the local connection 1-form of the principal bundle. In other words, the split of the Atiyah Lie algebroid is in one-to-one correspondence with the local gauge field.

\paragraph{Exterior algebra and the consistent splitting.}
The Atiyah Lie algebroid carries an exterior algebra
\begin{align}
\Omega^\bullet(A)=\bigoplus_{n=0}^{\mathrm{rank}A}\Omega^n(A)\,,
\end{align}
with Lie algebroid differential $\hat\td:\Omega^p(A)\to\Omega^{p+1}(A)$ defined by the Koszul formula (see Appendix~\ref{App:ALA}). More generally, given a vector bundle $F$ over $M$, we may also introduce the exterior algebra valued in $F$,
\begin{align}
\Omega^\bullet(A;F)=\bigoplus_{n=0}^{\mathrm{rank}A}\Omega^n(A;F)\,,
\end{align}
together with the corresponding Lie algebroid differential. This extension will be useful, for example, when $F$ describes charged matter fields transforming in a representation of the gauge algebra. In particular, the connection reform should be regarded as an $L$-valued 1-form $
\omega\in\Omega^1(A;L)$.

Once the algebroid is locally trivialized, the exterior algebra decomposes into a bicomplex according to the de~Rham form degree $p$ on $M$ and the degree $q$ in the Chevalley--Eilenberg algebra of $L$:
\begin{align}
\Omega^n(A)=\bigoplus_{p+q=n}\Omega^{(p,q)}(M,L)\,,
\qquad
\Omega^{(p,q)}(M,L)=\wedge^pT^*M\otimes \wedge^qL^*\,.
\end{align}
We refer to this as the \emph{consistent splitting}\footnote{This is to distinguish it from the \emph{covariant splitting}, where one writes $\Omega^n(A)=\bigoplus_{p+q=n}\Omega^{(p,q)}(H,V)$. The corresponding cohomologies in these two splittings lead to the consistent and covariant anomalies, respectively \cite{Jia:2023tki}.} of the exterior algebra. Physically, in this splitting $p$ counts the form degree on the base while $q$ counts the ghost number.

Let $\{\td x^\mu\}$ denote the basis of $T^*U_i$ dual to $\{\un\p_\mu\}$, and let $\{t^A\}$ be the basis of $L_i^*$ dual to $\{\un t_A\}$. Using the dual basis $\{\un E^{\un\alpha},\un E^{\un A}\}$:
\begin{align}
\un E^{\un\alpha}=\sigma^{\un\alpha}{}_{\mu}\td x^\mu\,,\qquad \un E^{\un A}=j^{\un A}{}_{A}(t^A-b^A_{i,\mu}\td x^\mu)\,.
\end{align}
We can expand any $\beta\in\Omega^1(A)$ as
\begin{align}
\beta&=\beta_{\un\alpha}\un E^{\un\alpha}+\beta_{\un A}\un E^{\un A}\\
&=(\beta_{\un\alpha}\sigma^{\un\alpha}{}_{\mu}-\beta_{\un A}j^{\un A}{}_{A}b^A_{i,\mu})\td x^\mu+\beta_{\un A}j^{\un A}{}_{A}t^A\nn\\
\label{betadecompo}
&=\Lambda_{i,\mu} \td x^\mu+m_{A}t^A\,,
\end{align}
where 
\begin{align}
\Lambda_{i,\mu}=(\tilde \Lambda_{i,\mu}-m_Ab^A_{i,\mu})\,,\qquad\tilde \Lambda_{i,\mu}=\beta_{\un\alpha}\sigma^{\un\alpha}{}_{\mu}\,,\qquad m_{A}=\beta_{\un A}j^{\un A}{}_{A}\,.
\end{align}
Therefore, in the consistent splitting, any $\beta\in\Omega^1(A)$ can be represented by a pair
\begin{align}
(\Lambda_i,m)\in\Omega^{(1,0)}(M,L)\oplus\Omega^{(0,1)}(M,L)\,.
\end{align}
This can be easily generalized to any $\beta\in\Omega^1(A,F)$. In particular, the connection reform $
\omega\in\Omega^1(A;L)$ can be written as
\begin{align}
\omega=(b^A_{i,\mu}\td x^\mu-t^A)\otimes \un t_A=b_i-\varpi\,,
\end{align}
where
\begin{align}
\varpi=t^A\otimes \un t_A
\end{align}
is the Maurer--Cartan form on the isotropy bundle $L$. This is the precise algebroid counterpart of the extended connection in the BRST formalism: $b_i$ plays the role of the gauge field with de~Rham degree one, and $\varpi$ plays the role of the ghost with ghost number one.

\paragraph{The Lie algebroid differential as $\td+\ts$.}
We now recall how the Lie algebroid differential resolves into the de~Rham and BRST operators. Let $\psi$ be a section of a vector bundle $F$ carrying a representation $v_F:L\to \End(F)$. Suppose $\{e_a\}$ is a basis of the sections on $F$. In the consistent splitting, the Lie algebroid differential acts as
\begin{align}
\hat\td\psi=\td\psi^a\otimes e_a+v_F(\un t_A)^a{}_bt^A\psi^b\otimes e_a\,.
\end{align}
Then we have
\begin{align}
\hat\td\psi=(\td+\ts)\psi
\end{align}
with the identification
\begin{align}
\ts\psi:=v_F(\un t_A)t^A\psi\,.
\end{align}
Moreover, for any $\beta\in\Omega^1(A;F)$ written as
\begin{align}
\beta=(\Lambda^a_\mu\td x^\mu+m_A^a t^A)\otimes e_a\,,
\end{align}
a direct computation gives
\begin{align}
\hat\td\beta
={}&
(\td\Lambda^a+v_F(\un t_A)^a{}_b\,t^A\wedge \Lambda^b)\otimes e_a \nn\\
&+
\Big(\td m_A^a\wedge t^A+\big(v_F(\un t_A)^a{}_b\,m_B^b-\tfrac12 m_C^a f^C{}_{AB}\big)t^A\wedge t^B\Big)\otimes e_a\\
={}&(\td+\ts)\Lambda_\mu^a\wedge \td x^\mu\otimes e_a+(\td+\ts)m_A^a\wedge t^A\otimes e_a\,,
\end{align}
once we identify
\begin{align} 
\ts\Lambda^a_\mu=v_F(\un t_A) ^a{}_bt^A\Lambda^b_\mu\,,\qquad \ts m_A^a=v_F(\un t_A) ^a{}_b m_B^bt^A-\frac{1}{2}m_C^af^C{}_{AB}t^A\,.
\end{align}
Thus, in the consistent splitting, $\hat\td$ can be interpreted as acting as $\td+\ts$, where the BRST operator $\ts$ is the Chevalley--Eilenberg differential acting along the isotropy directions.

In particular, when $\beta=\omega\in\Omega^1(A;L)$ is the connection reform, one obtains
\begin{align}
\hat\td\omega
=
(\td+\ts)(b_i^A\un t_A-\varpi^A\un t_A)\,,
\end{align}
with
\begin{align}
\ts b_i^A=\td\varpi^A+f^A{}_{BC}\,t^B\wedge b_i^C\,,
\qquad
\ts\varpi^A=\frac12 f^A{}_{BC}\,t^B\wedge t^C\,.
\end{align}

\paragraph{The Russian formula.}
The curvature of the connection reform can be defined as
\begin{align}
\label{curvaturereform}
\Omega=\hat\td\omega+\frac12[\omega,\omega]_L\,.
\end{align}
A basic geometric fact is that this curvature is horizontal. Substituting $\omega=b_i-\varpi$, one therefore finds
\begin{align}
\label{Omegahorizontal}
\Omega&=(\td+\ts)(b_i-\varpi)+\frac12[b_i-\varpi,b_i-\varpi]_L\nn\\
&=\td b_i+\frac12[b_i,b_i]_L\,.
\end{align}
Hence, the extended connection $b_i-\varpi$ has curvature equal to the ordinary gauge field strength
\begin{align}
F_i=\td b_i+\frac12[b_i,b_i]_L\,.
\end{align}
This gives precisely the Russian formula in the trivialized algebroid picture. In the Abelian case, which is the one relevant for the higher-form generalization studied later, the Lie bracket on $L$ vanishes and the formula reduces to
\begin{align}
\label{Russian1}
(\td+\ts)(b_i-\varpi)=F_i\,,
\qquad
F_i=\td b_i\,.
\end{align}

The trivialized Atiyah algebroid therefore furnishes a direct geometric realization of the BRST bicomplex. However, in order to generalize this construction to gerbes, it is more convenient to recast the same structure in \v Cech language. We now explain this \v Cech presentation.

\subsection{Atiyah Lie Algebroids in \v Cech Presentation}
\label{sec:ALACech}

From now on we will restrict to the case $G=U(1)$, since this is the case relevant for the higher-form generalization.

Given an open cover $\{U_i\}$ of $M$, a principal $U(1)$-bundle is specified by transition functions
\begin{align}
g_{ij}:U_{ij}\to U(1)
\end{align}
on double overlaps $U_{ij}=U_i\cap U_j$, satisfying the cocycle condition
\begin{align}
g_{ij}(x)g_{jk}(x)g_{ki}(x)=1\qquad \forall x\in U_{ijk}=U_i\cap U_j\cap U_k\,.
\end{align}
We denote the corresponding principal bundle by $P=\{g_{ij}\}$.

Locally, the Atiyah Lie algebroid is identified as
\begin{align}
A|_{U_i}\simeq TU_i\oplus L_i\,,
\qquad
L_i=L|_{U_i}\,.
\end{align}
Hence, a section over $U_i$ can be represented by a pair $(\uX,f^X_i)$, with $\uX\in\Gamma(TU_i)$ and $f_i^X\in\Omega^0(U_i,\mathfrak u(1))$.\footnote{Since in the $U(1)$ case $f_i$ has only a single component, we will omit the underscore and simply write $f_i$.} To recover the globally defined algebroid, we impose the patching condition
\begin{align}
\label{Apatch}
(\uX,f^X_i)\sim(\uX,f^X_j+i_X\td\ln g_{ij})\,,
\end{align}
where $\td\ln g_{ij}$ is shorthand for the Maurer--Cartan form $g_{ij}^{-1}\td g_{ij}$. This is the Abelian version of the usual transformation law \eqref{ALAintersect} for a section on the algebroid $A$. Then, we define ${\cal L}_{g_{ij}}$ to be a set where each element is a pair
\begin{align}
(\uX,\{f_i^X\})\,,
\end{align}
where $\uX\in\Gamma(TM)$ and $\{f_i^X\}$ is the collection of the local functions $f_i^X\in\Omega^0(U_i,\mathfrak u(1))$ for all $U_i$, satisfying
\begin{align}
\label{deltafi}
(\delta f^X)_{ij}=f_i^X-f_j^X=i_X\td\ln g_{ij}\,.
\end{align}
This can be viewed as the \v Cech presentation of the Atiyah Lie algebroid associated to the principal bundle $P=\{g_{ij}\}$. The Lie bracket on ${\cal L}_{g_{ij}}$ is induced from the local bracket \eqref{ALiebracket} on $TU_i\oplus L_i$. Since here the isotropy algebra is Abelian, the vertical bracket vanishes, and therefore
\begin{align}
\label{CechBracket}
[(\uX,\{f_i^X\}),(\uY,\{f_i^Y\})]_{{\cal L}_{g_{ij}}}
=
\big([\uX,\uY]_{TM},\{\lie_X f_i^Y-\lie_Y f_i^X\}\big)\,.
\end{align}
One readily checks that the right-hand side also satisfies \eqref{deltafi}, and so the bracket is globally well-defined. In this way, ${\cal L}_{g_{ij}}$ reproduces the Atiyah Lie algebroid in \v Cech language.

The corresponding exact sequence is now written at the level of sections:
\begin{equation}
\label{LgijSES}
\begin{tikzcd}
0
\arrow{r}
&
\Omega^0(M,\mathfrak u(1))
\arrow{r}{j}
&
{\cal L}_{g_{ij}}
\arrow{r}{\rho}
&
\Gamma(TM)
\arrow{r}
&
0\,.
\end{tikzcd}
\end{equation}
Since we are in the Abelian case, the isotropy bundle is trivial:
\begin{align}
L\cong M\times \mathfrak u(1)\,.
\end{align}
Nevertheless, the Atiyah Lie algebroid itself remains nontrivial through the transition functions of the principal bundle. The maps $\rho$ and $j$ are given by
\begin{align}
\rho(\uX,\{f_i^X\})=\uX\,,\qquad
j(\{f_i\})=(0,\{f_i\})\,,
\end{align}
where $\{f_i\}$ is the restriction of a global function $f\in\Omega^0(M,\mathfrak u(1))$ to the open sets $U_i$, so that $f_i=f_j$ on each overlap $U_{ij}$. Thus, the vertical part of ${\cal L}_{g_{ij}}$ consists precisely of those pairs with vanishing vector field.

Here ${\cal L}_{g_{ij}}$ should be understood as the \v Cech presentation of the \emph{sections} of the Atiyah Lie algebroid $A$. Accordingly, the exact sequence \eqref{LgijSES} is written at the level of sections rather than vector bundles. More intrinsically, the same construction may be formulated in terms of sheaves or categories. In particular, the categorical point of view makes clear that a Lie algebroid is a many-object analogue of a Lie algebra over $M$. Since our main goal here is to make the local BRST structure explicit, we will work directly with the corresponding \v Cech data.

\paragraph{Connection and local gauge field.}
A split of ${\cal L}_{g_{ij}}$ is specified by a horizontal lift of vector fields. In the \v Cech picture, this means that for each $\uX\in\Gamma(TM)$ we choose a distinguished lift
\begin{align}
\sigma(\uX)=(\uX,\{f_i^{X^H}\})\,,
\end{align}
where we define
\begin{align}
\label{fXH}
f_i^{X^H}=i_X b_i\,.
\end{align}
From \eqref{deltafi} we can see that the local 1-forms $b_i\in\Omega^1(U_i,\mathfrak u(1))$ satisfy
\begin{align}
\label{bpatch}
(\delta b)_{ij}=b_i-b_j=\td\ln g_{ij}\,.
\end{align}
Hence, the choice of split $\sigma$ is equivalent to the choice of a local gauge field $b_i$, namely a connection on the principal bundle.

Every section $(\uX,\{f_i^X\})$ can be decomposed into horizontal and vertical parts as follows:
\begin{align}
(\uX,\{f_i^X\})=(\uX,\{f_i^{X^H}\})+(0,\{\tilde f_i^X\})\,,
\end{align}
where $\tilde f_i^X\equiv f_i^X-f_i^{X^H}$. This defines the connection reform satisfying $\ker(\omega)=\im(\sigma)$ (note again that we follow the convention with minus sign):
\begin{align}
\omega(\uX,\{f_i^X\})=\{-\tilde f_i^X\}\,,
\end{align}
The maps $\rho$, $j$, $\sigma$ and $\omega$ therefore satisfy
\begin{align}
\rho(\uX,\{f_i^X\})&=\uX\,,\qquad
\sigma(\uX)=(\uX,\{f_i^{X^H}\})\,,\nn\\
\omega(\uX,\{f_i^X\})&=\{-\tilde f_i^X\}\,,\qquad
j(\{f_i\})=(0,\{f_i\})\,.
\end{align}
Moreover, it is easy to see that
\begin{align}
\rho\circ\sigma(\uX)&=\uX\,,\qquad
\omega\circ j(\{f_i\})=\{-f_i\}\,,\\
\sigma\circ\rho(\uX,\{f_i^X\})&=(\uX,\{f_i^{X^H}\})\,,\qquad
j\circ\omega(\uX,\{f_i^X\})=(0,\{-\tilde f_i^X\})\,.
\end{align}
We can see that $\rho\circ\sigma$ and $-\omega\circ j$ are identity maps, while $\sigma\circ\rho$ and $-j\circ\omega$ are the horizontal and vertical projections, respectively. Therefore, once the split is chosen, the exact sequence goes in both directions:
\begin{equation}
\label{LgijSESsplit}
\begin{tikzcd}
0
\arrow{r}
&
\Omega^0(M,\mathfrak u(1))
\arrow{r}{j}
\arrow[bend left]{l}
&
{\cal L}_{g_{ij}}
\arrow{r}{\rho}
\arrow[bend left]{l}{\omega}
&
\Gamma(TM)
\arrow{r}
\arrow[bend left]{l}{\sigma}
&
0\,.
\arrow[bend left]{l}
\end{tikzcd}
\end{equation}

\paragraph{Curvature and gauge field strength.}
Given two vector fields $\uX,\uY\in\Gamma(TM)$, the curvature of the split $\sigma$ is
\begin{align}
R^\sigma(\uX,\uY)
&=
[\sigma(\uX),\sigma(\uY)]_{{\cal L}_{g_{ij}}}-\sigma([\uX,\uY]_{TM})\nn\\
&=
[(\uX,\{f_i^{X^H}\}),(\uY,\{f_i^{Y^H}\})]_{{\cal L}_{g_{ij}}}
-([\uX,\uY],\{f_i^{[X,Y]^H}\})\nn\\
&=
\big(0,\{f_i^{[X^H,Y^H]}-f_i^{[X,Y]^H}\}\big)\,.
\end{align}
Using \eqref{fXH}, we can compute that
\begin{align}
f_i^{[X^H,Y^H]}-f_i^{[X,Y]^H}=
\lie_X i_Y b_i-\lie_Y i_X b_i-i_{[X,Y]}b_i=i_Yi_XF_i\,,
\end{align}
where $F_i=\td b_i$. Hence,
\begin{align}
\label{RsigmaCech}
R^\sigma(\uX,\uY)=\big(0,\{i_Yi_XF_i\}\big)\,.
\end{align}

One may similarly compute the curvature of the connection reform $\omega$ on horizontal sections:
\begin{align}
&R^\omega((\uX,\{f_i^{X^H}\}),(\uY,\{f_i^{Y^H}\}))\nn\\
={}&
[\omega(\uX,\{f_i^{X^H}\}),\omega(\uY,\{f_i^{Y^H}\})]_L
-\omega([(\uX,\{f_i^{X^H}\}),(\uY,\{f_i^{Y^H}\})]_{{\cal L}_{g_{ij}}})\,.
\end{align}
Since the isotropy algebra is Abelian, the first term vanishes, and therefore
\begin{align}
R^\omega((\uX,\{f_i^{X^H}\}),(\uY,\{f_i^{Y^H}\}))
=\big\{f_i^{[X,Y]^H}-f_i^{[X^H,Y^H]}\big\}=\{-i_Yi_XF_i\}\,.
\end{align}
Thus,
\begin{align}
\label{curvatureeq}
R^\sigma(\uX,\uY)
=-j\Big(R^\omega((\uX,\{f_i^{X^H}\}),(\uY,\{f_i^{Y^H}\}))\Big)
=\big(0,\{i_Yi_XF_i\}\big)\,.
\end{align}
In this way, the curvatures defined from $\sigma$, $\omega$, and the local gauge field $b_i$ all coincide and agree with the curvature reform \eqref{curvaturereform}. In other words, all these notions of curvature encode the same geometric obstruction on the algebroid.

Finally, from \eqref{bpatch} we have
\begin{align}
(\delta F)_{ij}=F_i-F_j=\td(\delta b)_{ij}=\td^2\ln g_{ij}=0\,,
\end{align}
and so the local 2-forms $F_i$ glue to a globally defined 2-form
\begin{align}
\label{globalF}
F\in\Omega^2(M)\,.
\end{align}
This is precisely the curvature of the principal $U(1)$-connection.

\subsection{Comparison with the \v Cech--de~Rham Complex}
\label{sec:PhiMap}
We now explain how the \v Cech presentation ${\cal L}_{g_{ij}}$ is related to the bicomplex of the trivialized Atiyah Lie algebroid reviewed above. In the Abelian pure gauge sector, we show that the \v Cech--de~Rham description established in \cite{Alvarez:1984es} agrees with the BRST bicomplex of the trivialized Atiyah Lie algebroid. In this way, the map $\Phi$ constructed below provides the bridge between the two descriptions. This sets up the stage for the later generalization to the gerbe case, since the \v Cech formulation is expressed directly in terms of local descent data and therefore continues to make sense when no ordinary bundle description is available.

In a consistent splitting, the exterior algebra of the Atiyah Lie algebroid takes the form
\begin{align}
\Omega^n(A)=\bigoplus_{p+q=n}\Omega^{(p,q)}(M,L)\,,
\end{align}
and the Lie algebroid differential $\hat\td$ decomposes as $\hat\td=\td+\ts$, with
\begin{align}
\td&:\Omega^{(p,q)}(M,L)\to \Omega^{(p+1,q)}(M,L)\,,\\
\ts&:\Omega^{(p,q)}(M,L)\to \Omega^{(p,q+1)}(M,L)\,.
\end{align}
On the other hand, given an open cover ${\cal U}=\{U_i\}$, the \v Cech--de~Rham bicomplex is
\begin{align}
C^\bullet({\cal U},\Omega^\bullet(M))\,,
\end{align}
with differentials
\begin{align}
\td&:C^q({\cal U},\Omega^p(M))\to C^q({\cal U},\Omega^{p+1}(M))\,,\\
\delta&:C^q({\cal U},\Omega^p(M))\to C^{q+1}({\cal U},\Omega^p(M))\,.
\end{align}
In the Abelian case there is a natural comparison map
\begin{align}
\Phi:\Omega^{(p,q)}(M,L)\longrightarrow C^q({\cal U},\Omega^p(M))
\end{align}
defined on generators by
\begin{align}
\Phi(\td x^\mu)=\td x^\mu,\qquad
\Phi(t^*)=\varpi_{ij}\equiv\ln g_{ij}\,.
\end{align}
Thus, the generator of Chevalley--Eilenberg degree one is mapped to the \v Cech 1-cochain determined by the transition functions.

The basic compatibility statement is that, in the pure gauge sector, $\Phi$ intertwines the two bicomplex differentials:
\begin{align}
\label{Phichainmap}
\Phi\circ(\td+\ts)=(\td+\delta)\circ\Phi\,.
\end{align}
Since the graded algebra $\Omega^\bullet(A)$ is generated by degree 0 and degree 1 elements, and both $\hat\td=\td+\ts$ and $\td+\delta$ act as degree-one derivations, it suffices to verify \eqref{Phichainmap} on $\Omega^0(A)$ and $\Omega^1(A)$.

For a scalar $\beta\in\Omega^0(A)$, the statement is immediate. We now consider a degree-one element
$\beta\in \Omega^1(A)$, which in the consistent splitting can be represented by
\begin{align}
\beta=(\Lambda,m)\in \Omega^{(1,0)}(M,L)\oplus \Omega^{(0,1)}(M,L)\,.
\end{align}
Recalling the decomposition \eqref{betadecompo}, under the map $\Phi$ one has
\begin{align}
\Phi(\beta)=\Lambda_i+m\,\varpi_{ij}
=(\tilde\Lambda_i-mb_i)+m\,\varpi_{ij}\,.
\end{align}
Since the isotropy algebra is Abelian, the Lie algebroid differential reduces to
\begin{align}
(\td+\ts)\beta=\td\Lambda+\td m\wedge t^*\,.
\end{align}
Applying $\Phi$ gives
\begin{align}
\Phi((\td+\ts)\beta)=\td\Lambda_i+\td m\,\varpi_{ij}\,.
\end{align}
On the other hand, the \v Cech--de~Rham differential acts by
\begin{align}
(\td+\delta)\Phi(\beta)
&=(\td+\delta)(\Lambda_i+m\,\varpi_{ij})\nn\\
&=\td\Lambda_i+\delta\Lambda_i+\td m\,\varpi_{ij}+m\,\td\varpi_{ij}+m\,(\delta\varpi)_{ijk}\,.
\end{align}
Using $(\delta\varpi)_{ijk}=0$ and $(\delta b)_{ij}=\td\varpi_{ij}$, we obtain
\begin{align}
(\td+\delta)\Phi(\beta)=\td\Lambda_i+\td m\,\varpi_{ij}
=\Phi((\td+\ts)\beta)\,,
\end{align}
and hence \eqref{Phichainmap} holds for degree-one elements in the pure gauge sector.

The comparison map $\Phi$ is to be understood here in the Abelian pure gauge sector, which is the only case needed for the Russian formula and the descent equations discussed in this paper. In that sector, the \v Cech--de~Rham presentation is built entirely from the transition data $g_{ij}$, and the identification of the Chevalley--Eilenberg degree-one generator $t^*$ with the \v Cech 1-cochain $\varpi_{ij}=\ln g_{ij}$ gives the expected correspondence between the BRST bicomplex and the \v Cech--de~Rham bicomplex.

In the presence of charged matter fields, however, one needs to consider cochains valued in a vector bundle $F$, and the naive extension of $\Phi$ is no longer a strict cochain map for $\beta\in\Omega^1(A;F)$. The reason is that the twisting coming from the $U(1)$-representation produces additional overlap terms which are absent in the pure gauge case. A satisfactory treatment of such $F$-valued cochains would require a more refined comparison morphism \cite{crainic2003differentiable,cabrera2017van}, and we shall not pursue this here.

Therefore, in the Abelian pure gauge sector, the map $\Phi$ provides the desired chain-level comparison between the BRST bicomplex of the trivialized Atiyah Lie algebroid and the \v Cech--de~Rham bicomplex of its \v Cech presentation ${\cal L}_{g_{ij}}$. In this sense, the \v Cech differential $\delta$ may be regarded as the \v Cech realization of the BRST differential $\ts$.
\subsection{Russian Formula and the BRST Complex}

Physically, one identifies $b_i$ as the local gauge field and $F_i=\td b_i$ as the local gauge field strength. Recall from \eqref{globalF} that the local 2-forms $F_i$ glue to a globally defined field strength $F\in\Omega^2(M)$. We may now reexpress the horizontality of the curvature of the connection reform $\omega=b-\varpi$ in the \v Cech language as
\begin{align}
\label{Russianf1}
\boxed{(\td+\delta)(b-\varpi)=F}\,.
\end{align}
This is precisely the Russian formula in the \v Cech--de~Rham bicomplex.

Eq.~\eqref{Russianf1} is the image under the comparison map $\Phi$ of the corresponding statement \eqref{Russian1} in the trivialized Atiyah Lie algebroid. Through the chain map \eqref{Phichainmap}, the BRST differential $\ts$ is represented in the \v Cech language by $\delta$, while the Chevalley--Eilenberg generator $t^*$ is realized at the level of local cocycle data by the \v Cech 1-cochain $\varpi_{ij}=\ln g_{ij}$. Thus, the \v Cech--de~Rham bicomplex provides a local realization of the BRST complex.

We identify ghost number with \v Cech degree in the \v Cech--de~Rham bicomplex. Under the comparison map, the \v Cech differential $\delta$ realizes the BRST operator $\ts$, while $-\varpi$ plays the role of the ghost field. At first sight, one might recognize $\delta$ as the classical gauge variation and $\ln g_{ij}$ as the transformation parameter. The fact that they remain encoded in the structure designed for the quantized theory is nothing mystical: both $(\delta,\varpi)$ and $(\ts,c)$ arise from the same underlying cohomological structure associated with gauge symmetry, and that structure is already present at the classical level. In the BRST formalism, one resolves the quotient by gauge orbits by adjoining ghost variables together with a nilpotent differential. In the Atiyah Lie algebroid, this resolution is already built into the exact sequence \eqref{LgijSES} under trivialization, and the geometric fact that the algebroid curvature $\Omega$ is horizontal is precisely what gives rise to the Russian formula \eqref{Russianf1}. Thus, in the Abelian case, $\delta$ plays for gauge directions a role analogous to that of the de~Rham differential $\td$ for spacetime directions: $\td$ differentiates along $TM$, while $\delta$ probes the gauge directions encoded by the transition functions. The cocycle $\varpi_{ij}=\ln g_{ij}$ is the corresponding local datum, and in this sense it may be regarded as the classical antecedent of the ghost field in the BRST formalism.

\section{Lie 2-Algebroids Derived from Gerbes}
\label{sec:Lie2Alg}

In the previous section we reviewed the Atiyah Lie algebroid and formulated its \v Cech presentation ${\cal L}_{g_{ij}}$. The point of that discussion was not merely to recast the ordinary BRST complex in another language, but to exhibit the structure in a form that admits a direct higher analogue. Physically, this higher analogue is required because the background gauge field for a 1-form symmetry is not an ordinary connection on a principal bundle, but a gerbe with connective structure. In the case of a gerbe, there is no ordinary vector bundle whose sections directly play the role of the Atiyah Lie algebroid. Our strategy is therefore to work directly in the \v Cech picture, where the analogue of ${\cal L}_{g_{ij}}$ arises naturally from the gerbe cocycle itself.

The basic claim of this section is that the \v Cech data of a $U(1)$-gerbe determines a higher infinitesimal symmetry object ${\cal L}_{g_{ijk}}$, which should be regarded as the Lie 2-algebroid associated to the gerbe. For our purposes, it is sufficient to work with this object directly in terms of local data and their patching relations. More intrinsically, one may formulate the constructions below in terms of sheaves or 2-categories, as developed in \cite{collier2011}. Since our main aim is to make the local BV--BRST structure explicit, we will not emphasize that language in what follows.

\subsection{$U(1)$-Gerbes and the \v Cech Presentation of Lie 2-Algebroids}

Fix an open cover $\{U_i\}$ of the base manifold $M$. A $U(1)$-gerbe $Q$ over $M$ is specified by a \v Cech 2-cocycle\footnote{In the mathematical literature there are several models of gerbes \cite{Brylinski1993,murray1996bundle,Chatterjee1998,Hitchin:1999fh,Murray:2007ps,Schweigert2011}. Here we use the Hitchin--Chatterjee description \cite{Chatterjee1998,Hitchin:1999fh} (equivalent to Murray's bundle gerbes \cite{murray1996bundle,Murray:2007ps}), which is particularly convenient in \v Cech language.}
\begin{equation}
Q=\{g_{ijk}:U_{ijk}\to U(1)\}
\end{equation}
defined on triple overlaps $U_{ijk}=U_i\cap U_j\cap U_k$, and satisfying on quadruple overlaps $U_{ijkl}$ that
\begin{equation}
g_{ijk}g_{ikl}=g_{ijl}g_{jkl}\,.
\end{equation}
In \v Cech notation this is simply $\delta g=1$. Different choices of $\{g_{ijk}\}$ related by a \v Cech coboundary $g\mapsto g\,\delta h$ describe isomorphic gerbes. The associated cohomology class $[g]\in H^2(M,U(1))\cong H^3(M,\mathbb Z)$ is called the \emph{Dixmier--Douady (DD) class}.

In the last section, we have seen that a lift of a vector field $\uX$ to the Atiyah Lie algebroid can be described in the \v Cech language by a collection of local functions $\{f_i^X\}$ satisfying $(\delta f^X)_{ij}=i_X\td\ln g_{ij}$. In the case of gerbes, it is natural to obtain the higher analogue of ${\cal L}_{g_{ij}}$ for principal bundles by shifting the \v Cech degree up by one. We therefore define ${\cal L}_{g_{ijk}}$ to be the collection of pairs
\begin{equation}
(\uX,\{f_{ij}^X\})\,,
\end{equation}
where $\uX\in\Gamma(TM)$ and the local functions $f_{ij}^X\in \Omega^0(U_{ij},\mathfrak u(1))$
satisfy the cocycle condition \cite{collier2011}
\begin{equation}
\label{deltafij}
(\delta f^X)_{ijk}=f_{ij}^X+f_{jk}^X-f_{ik}^X=i_X\td\ln g_{ijk}\,.
\end{equation}
This is the direct higher analogue of the defining relation \eqref{deltafi} for ${\cal L}_{g_{ij}}$, with the gerbe cocycle $g_{ijk}$ replacing the principal bundle cocycle $g_{ij}$.

We may think of an element $(\uX,\{f_{ij}^X\})$ as a lift of the vector field $\uX$. Since the data now lives one \v Cech degree higher, such lifts are determined only up to a further local choice. Concretely, given another collection $\{f_{ij}^{\prime X}\}$ satisfying \eqref{deltafij}, the two lifts are related if there exist local functions $u_i\in \Omega^0(U_i,\mathfrak u(1))$ such that
\begin{equation}
\label{uimorphism}
f_{ij}^{\prime X}=f_{ij}^X+u_j-u_i\,.
\end{equation}
Since $\delta^2=0$, the transformation \eqref{uimorphism} preserves the cocycle condition \eqref{deltafij}. Conversely, if both $\{f_{ij}^X\}$ and $\{f_{ij}^{\prime X}\}$ satisfy \eqref{deltafij}, then their difference is a \v Cech 1-cocycle,
\begin{equation}
\delta(f^{\prime X}-f^X)=0\,,
\end{equation}
and hence on a good cover is locally exact. Therefore, any two lifts of the same vector field are related by a collection $\{u_i\}$ of the form \eqref{uimorphism}.\footnote{In \cite{collier2011}, $(\uX,\{f_{ij}^X\})$ is introduced an object of the category ${\cal L}_{g_{ijk}}$, and $\{u_i\}$ is the morphism between $(\uX,\{f_{ij}^X\})$ and $(\uX,\{f_{ij}^{\prime X}\})$. In the 2-categorical point of view, one may add an additional layer by regarding the points of $M$ as objects, and $(\uX,\{f_{ij}^X\})$ as a 1-morphism over the same point and $\{u_i\}$ as a 2-morphism.}

We can consider the anchor map as follows:
\begin{equation}
\rho_2(\uX,\{f_{ij}^X\})=\uX\,.
\end{equation}
Then, we may think of ${\cal L}_{g_{ijk}}$ as fitting into a higher analogue of the exact sequence \eqref{LgijSES}:
\begin{equation}
\label{LgijkSES}
\begin{tikzcd}
0
\arrow{r}
&
C^1({\cal U},\Omega^0(M,\mathfrak u(1)))
\arrow{r}{j_2}
&
{\cal L}_{g_{ijk}}
\arrow{r}{\rho_2}
&
\Gamma(TM)
\arrow{r}
&
0\,.
\end{tikzcd}
\end{equation}
where $j_2$ acts by
\begin{equation}
j_2(\{f_{ij}\})=(0,\{f_{ij}\})\,.
\end{equation}
The kernel of the anchor map consists of those pairs with vanishing vector field; that is, collections $\{f_{ij}\}$ satisfying $(\delta f)_{ijk}=0$, so that $\{f_{ij}\}$ defines a \v Cech $1$-cocycle and is therefore locally exact. 

The local bracket on ${\cal L}_{g_{ijk}}$ is defined in the same way as \eqref{CechBracket} for the ordinary symmetry case:
\begin{equation}
\label{L2bracket}
[(\uX,\{f_{ij}^X\}),(\uY,\{f_{ij}^Y\})]_{{\cal L}_{g_{ijk}}}
=
\big([\uX,\uY]_{TM},\{\lie_X f_{ij}^Y-\lie_Y f_{ij}^X\}\big)\,.
\end{equation}
One can readily check that the bracketed object again satisfies the cocycle condition \eqref{deltafij}, since we can find by using the Cartan identity that
\begin{align}
(\delta(\lie_X f^Y-\lie_Y f^X))_{ijk}
&=
\lie_X(\delta f^Y)_{ijk}-\lie_Y(\delta f^X)_{ijk}\nn\\
&=
\lie_X(i_Y\td\ln g_{ijk})-\lie_Y(i_X\td\ln g_{ijk})\nn\\
&=
i_{[X,Y]}\td\ln g_{ijk}\,.
\end{align}
In this way, the algebraic structure carried by ${\cal L}_{g_{ij}}$ extends directly to the higher counterpart ${\cal L}_{g_{ijk}}$. For our purposes, this is the sense in which ${\cal L}_{g_{ijk}}$ should be regarded as a Lie 2-algebroid: it is a 2-term local symmetry object built from the gerbe cocycle, equipped with an anchor to vector fields and a bracket compatible with the higher cocycle condition, even though we do not have a vector bundle. Therefore, we will refer to ${\cal L}_{g_{ijk}}$ as the \v Cech presentation of a Lie 2-algebroid. We will see that this is precisely the structure needed to organize the infinitesimal symmetries relevant to the BV--BRST complex. 

\paragraph{Split and connective structure.}
A $U(1)$-gerbe with \emph{connective structure} is described in addition by a collection of 1-forms $C=\{C_{ij}\in\Omega^1(U_{ij})\}$ defined on double overlaps, and satisfying on triple overlaps
\begin{equation}
(\delta C)_{ijk}=C_{ij}+C_{jk}-C_{ik}=\td\ln g_{ijk}\,.
\end{equation}
Thus, $C_{ij}$ is a \v Cech 1-cochain whose coboundary reproduces the \v Cech 2-cocycle $\td\ln g_{ijk}$ at the level of 1-forms, just as a connection $b_i$ has coboundary $\td\ln g_{ij}$ in the principal bundle case. 

We now show that the split of the higher symmetry object ${\cal L}_{g_{ijk}}$ is precisely the connective structure of the gerbe. This is the direct analogue of the fact that a split of ${\cal L}_{g_{ij}}$ is equivalent to the local gauge field $b_i$.

A split of ${\cal L}_{g_{ijk}}$ means that, for each vector field $\uX\in\Gamma(TM)$, we choose a distinguished lift through a map $\sigma_2:\Gamma(TM)\to{\cal L}_{g_{ijk}}$:
\begin{equation}
\sigma_2(\uX)=(\uX,\{f_{ij}^{X^H}\})\,.
\end{equation}
The natural higher analogue of the Atiyah formula $f_i^{X^H}=i_Xb_i$ is
\begin{equation}
\label{fXHC}
f_{ij}^{X^H}=i_X C_{ij}\,,
\end{equation}
where $C_{ij}\in\Omega^1(U_{ij},\mathfrak u(1))$ is a collection of 1-forms on double overlaps. Substituting \eqref{fXHC} into the cocycle condition \eqref{deltafij}, we obtain
\begin{equation}
\label{deltaC}
(\delta C)_{ijk}=C_{ij}+C_{jk}-C_{ik}=\td\ln g_{ijk}\,.
\end{equation}
which is precisely the defining condition for a connective structure on the gerbe. Therefore, a split of ${\cal L}_{g_{ijk}}$ is equivalent to the choice of connective structure on the gerbe. 

Once a split is chosen, every element of ${\cal L}_{g_{ijk}}$ decomposes as
\begin{equation}
(\uX,\{f_{ij}^X\})
=
(\uX,\{f_{ij}^{X^H}\})+(0,\{f_{ij}^X-f_{ij}^{X^H}\})\,.
\end{equation}
Hence, we can introduce the higher connection reform
\begin{equation}
\omega_2(\uX,\{f_{ij}^X\})=\{f_{ij}^{X^H}-f_{ij}^X\}\,,
\end{equation}
which satisfies $\ker(\omega_2)=\im(\sigma_2)$. Then, in exact analogy with the ordinary symmetry case, we have the following exact sequences in both directions:
\begin{equation}
\label{LgijkSESsplit}
\begin{tikzcd}
0
\arrow{r}
&
C^1({\cal U},\Omega^0(M,\mathfrak u(1)))
\arrow{r}{j_2}
\arrow[bend left]{l}
&
{\cal L}_{g_{ijk}}
\arrow{r}{\rho_2}
\arrow[bend left]{l}{\omega_2}
&
\Gamma(TM)
\arrow{r}
\arrow[bend left]{l}{\sigma_2}
&
0\,.
\arrow[bend left]{l}
\end{tikzcd}
\end{equation}
The maps $\rho_2$, $j_2$, $\sigma_2$ and $\omega_2$ satisfy
\begin{align}
\rho_2\circ\sigma_2(\un X)&=\un X\,,\qquad \sigma_2\circ\rho_2(\un X,\{f^X_{ij}\})=(\un X,\{f^{X^H}_{ij}\})\,,\\ 
\omega_2\circ j_2(\{f_{ij}\})&=\{f_{ij}\}\,,\qquad j_2\circ\omega_2(\un X,\{f^X_{ij}\})=(0,\{f^{X^H}_{ij}-f^X_{ij}\})\,.
\end{align}

\paragraph{Curvature of the split.}
We next compute the obstruction for the split $\sigma_2$ to preserve the bracket. Given two vector fields $\uX,\uY\in\Gamma(TM)$, the curvature of $\sigma_2$ is defined as
\begin{align}
R^{\sigma_2}(\uX,\uY)
&=
[\sigma_2(\uX),\sigma_2(\uY)]_{{\cal L}_{g_{ijk}}}
-\sigma_2([\uX,\uY]_{TM})\nn\\
&=
\big(0,\{f_{ij}^{[X^H,Y^H]}-f_{ij}^{[X,Y]^H}\}\big)\,.
\end{align}
Using \eqref{fXHC} and the Cartan identity, we find
\begin{align}
f_{ij}^{[X^H,Y^H]}-f_{ij}^{[X,Y]^H}
&=
\lie_X i_Y C_{ij}-\lie_Y i_X C_{ij}-i_{[X,Y]}C_{ij}\nn\\
&=
i_Yi_X\td C_{ij}\,,
\end{align}
and thus
\begin{equation}
\label{Rsigma2}
R^{\sigma_2}(\uX,\uY)=\big(0,\{i_Yi_X\td C_{ij}\}\big)\,.
\end{equation}
Equivalently, if one introduces the higher connection reform
\begin{equation}
\omega_2(\uX,\{f_{ij}^X\})=\{f_{ij}^{X^H}-f_{ij}^X\}\,,
\end{equation}
then its curvature is represented by the same local data,
\begin{equation}
R^{\omega_2}((\uX,\{f_{ij}^{X^H}\}),(\uY,\{f_{ij}^{Y^H}\}))
=\{-i_Yi_X\td C_{ij}\}\,,
\end{equation}
since the bracket on the Abelian vertical part vanishes. Thus, the curvatures of the split $\sigma_2$ and the connection form $\omega_2$ are encoded by the same \v Cech 1-cochain $\{i_Yi_X\td C_{ij}\}$, exactly as in \eqref{curvatureeq} for ordinary gauge symmetry.

At present we restrict attention to the higher gauge sector encoded by ${\cal L}_{g_{ijk}}$. A more complete higher gauge theory would presumably require an analogue of charged matter fields. Since gerbes are naturally associated with extended objects rather than point particles, such fields should be related to string-like charged objects. Correspondingly, one expects that this would involve an appropriate notion of higher representation of the Lie 2-algebroid ${\cal L}_{g_{ijk}}$. We do not develop this here, and leave it as a possible direction for extending the present construction beyond the pure higher gauge sector. 

\subsection{The \v Cech--de~Rham Complex and a Partial Higher Russian Formula}

For the Atiyah Lie algebroid, we began with the algebroid differential $\hat\td$, then passed through a local trivialization to the BRST form $\td+\ts$, and finally translated this to the \v Cech--de~Rham differential $\td+\delta$. In the gerbe case there is no natural vector bundle whose exterior algebra directly supplies such a differential. The higher symmetry structure is therefore most naturally described directly in the total \v Cech--de~Rham bicomplex.

In the \v Cech presentation, the algebra is exactly parallel to the ordinary symmetry case. There, a split determined by the local connection data $b_i$ led to the relation \eqref{Russianf1}; here, the split is determined by the connective structure $C_{ij}$. Denoting
\begin{equation}
c_{ijk}\equiv\ln g_{ijk}\,,
\end{equation}
we immediately obtain the following relation:
\begin{equation}
\label{higher-descent-Cc}
(\td+\delta)(C-c)=\td C\,,
\end{equation}
where $C=\{C_{ij}\}$ is viewed as a \v Cech 1-cochain of 1-forms, while $c=\{c_{ijk}\}$ is a \v Cech 2-cochain of 0-forms. Thus, the relation \eqref{higher-descent-Cc} is a direct higher analogue of the Russian formula \eqref{Russianf1} for ordinary gauge symmetry, with the connective structure $C_{ij}$ replacing the local gauge potential $b_i$.

Applying $\td$ to $\delta C=\td c$ yields $\delta(\td C)_{ijk}=0$, and thus $\{\td C_{ij}\}$ is a \v Cech 1-cocycle valued in $\Omega^2(M)$. Hence, one may choose a collection of local 2-forms $B_i\in\Omega^2(U_i,\mathfrak u(1))$ such that
\begin{equation}
\label{deltaB}
(\delta B)_{ij}=B_i-B_j=\td C_{ij}\,.
\end{equation}
Then, the relation \eqref{higher-descent-Cc} can be written as
\begin{equation}
\label{higher-descent-CcB}
(\td+\delta)(C-c)=\delta B\,.
\end{equation}
Given a gerbe with connective structure, such a local 2-form $B_i$ is called a \emph{curving} of the connective structure.

There is an important difference here from the ordinary gauge symmetry, where the curvature of the split lands directly in the vertical part of the algebroid. In the gerbe case, the curvature of the split \eqref{Rsigma2} may be written as
\begin{equation}
R^{\sigma_2}(\uX,\uY)=\big(0,\{f_{ij}^{[X^H,Y^H]}-f_{ij}^{[X,Y]^H}\}\big)=\big(0,\{i_Yi_X(B_i-B_j)\}\big)\,.
\end{equation}
Therefore, the transition \eqref{uimorphism} between the lift of the bracket and the bracket of the lifts is represented by
\begin{equation}
u_i=i_Yi_XB_i\,.
\end{equation}
In other words, the collection $\{B_i\}$ does not define the split of ${\cal L}_{g_{ijk}}$ itself, but controls the failure of the split to preserve the bracket.

However, the relation \eqref{higher-descent-CcB} is not yet the higher Russian formula itself. At this stage, the local 2-forms $B_i$ have only appeared as the curving associated to the split of the higher symmetry object ${\cal L}_{g_{ijk}}$. In the next section we will see that these same 2-forms define the split of the associated exact Courant algebroid. Once that interpretation is in place, we will obtain the full higher Russian formula and the geometric meaning of the BV--BRST tower becomes manifest.

\section{Exact Courant Algebroids Derived from Gerbes}
\label{sec:Courant}

In the previous section we introduced the \v Cech Lie 2-algebroid ${\cal L}_{g_{ijk}}$ associated to a $U(1)$-gerbe and showed that its split is precisely the connective structure $C_{ij}$. We also found that the local 2-forms $B_i$ appear naturally in the curvature data of that split. We now explain that these same 2-forms admit a second, equally natural geometric interpretation: they define the split of an exact Courant algebroid associated to the gerbe. This is the higher analogue of the role played by the local gauge field in the ordinary symmetry case. For a brief review of the basics of Courant algebroids, see Appendix~\ref{App:Courant}.

\subsection{Exact Courant Algebroids in \v Cech Language}

An exact Courant algebroid is a vector bundle $E\to M$ equipped with a Courant bracket $[\cdot,\cdot]_E$, a nondegenerate symmetric pairing $\langle\cdot,\cdot\rangle_E$, and an anchor map $\rho_E:E\to TM$, fitting into the exact sequence 
\begin{equation}
\label{Courant}
\begin{tikzcd}
0
\arrow{r}
&
T^*M
\arrow{r}{j_E}
&
E
\arrow{r}{\rho_E}
&
TM
\arrow{r}
&
0\,.
\end{tikzcd}
\end{equation}
For an exact Courant algebroid, however, the symmetric pairing determines the inclusion canonically: for each $\alpha\in T^*M$, the element $j_E(\alpha)\in E$ is characterized by
\begin{equation}
\langle\rho_E^*(\alpha),e\rangle_E=\alpha(\rho_E(e))
\qquad
\forall \alpha\in T^*M,\ e\in E\,.
\end{equation}
Thus, the inclusion $j_E$ is in fact precisely the adjoint map $\rho_E^*$.

As in the ordinary symmetry case, we can also work in the \v Cech presentation. Let $\{U_i\}$ be an open cover of $M$. On each patch $U_i$, an exact Courant algebroid is locally trivialized as
\begin{equation}
E|_{U_i}\cong TU_i\oplus T^*U_i\,.
\end{equation}
A local section is therefore represented by a pair $(\uX,a_i)$, with $\uX\in\Gamma(TU_i)$ and $a_i\in\Omega^1(U_i)$. To glue these local models into a global exact Courant algebroid, one uses closed 2-forms ${\cal B}_{ij}\in\Omega^2(U_{ij})$ satisfying
\begin{equation}
\label{Btriple}
(\delta {\cal B})_{ijk}={\cal B}_{ij}+{\cal B}_{jk}-{\cal B}_{ik}=0\,,
\qquad
\td{\cal B}_{ij}=0\,,
\end{equation}
and identifies on overlaps
\begin{equation}
\label{CApatchtrans}
(\uX,a_i)\sim (\uX,a_j+i_X{\cal B}_{ij})\,.
\end{equation}

We therefore define ${\cal L}_{{\cal B}_{ij}}$ to be the collection of pairs
\begin{equation}
(\uX,\{a_i^X\})\,,
\end{equation}
where $\uX\in\Gamma(TM)$ and the local 1-forms $a_i^X\in\Omega^1(U_i)$ satisfy $a_i^X=a_j^X+i_X{\cal B}_{ij}$. This is the \v Cech presentation of the exact Courant algebroid associated to the gluing data $\{{\cal B}_{ij}\}$.

At the level of sections, we obtain the exact sequence
\begin{equation}
\label{LCijSES}
\begin{tikzcd}
0
\arrow{r}
&
\Omega^1(M)
\arrow{r}{\rho_E^*}
&
{\cal L}_{{\cal B}_{ij}}
\arrow{r}{\rho_E}
&
\Gamma(TM)
\arrow{r}
&
0\,,
\end{tikzcd}
\end{equation}
where
\begin{equation}
\rho_E(\uX,\{a_i^X\})=\uX\,,
\qquad
\rho_E^*(a)=(0,\{a\})\,.
\end{equation}

The local Courant bracket on $TU_i\oplus T^*U_i$ induces a bracket on ${\cal L}_{{\cal B}_{ij}}$ given by
\begin{align}
\label{CechCourantBracket}
[(\uX,\{a_i^X\}),(\uY,\{a_i^Y\})]_{{\cal L}_{{\cal B}_{ij}}}=\Big(
[\uX,\uY]_{TM},\{\lie_X a_i^Y-\lie_Y a_i^X-\frac12\td(i_Xa_i^Y-i_Ya_i^X)\}\Big)\,,
\end{align}
and the pairing is given by
\begin{equation}
\label{pairing-local}
\langle(\uX,\{a_i^X\}),(\uY,\{a_i^Y\})\rangle_{{\cal L}_{{\cal B}_{ij}}}=\{i_Xa_i^Y+i_Ya_i^X\}\,.
\end{equation}
Using \eqref{Btriple}, one can check that both the bracket and the pairing are compatible with the identification \eqref{CApatchtrans}, and thus are globally well-defined.

\paragraph{Split, isotropic split, and curvature.}
A split of the exact sequence \eqref{LCijSES} is a map $\sigma_E:\Gamma(TM)\to {\cal L}_{{\cal B}_{ij}}$ such that
\begin{equation}
\sigma_E(\un X)=(\un X,\{a_i^{X^H}\})\,.
\end{equation}
Equivalently, we may define a connection reform $\omega_E:{\cal L}_{{\cal B}_{ij}}\to \Omega^1(M)$ with
\begin{equation}
j_E(\{a_i\})=(0,\{a_i\})\,,
\end{equation}
which satisfies $\ker(\omega_E)=\im(\sigma_E)$. The maps $\rho_E$, $j_E$, $\sigma_E$ and $\omega_E$ satisfy
\begin{align}
\rho_E\circ\sigma_E(\un X)&=\un X\,,\qquad\sigma_E\circ\rho_E((\un X,\{a_i^{X}\}))=(\un X,\{a_i^{X^H}\})\,,\\
\omega_E\circ j_E(\{a_i\})&=\{a_i\}\,,\qquad j_E\circ\omega_E((\un X,\{a_i^{X}\}))=(0,\{a_i^{X}-a_i^{X^H}\})\,.
\end{align}

Note that unlike $j_E$ and $\rho_E$, $\sigma_E$ is not automatically the pullback of $\omega_E$ since the image of $\omega_E$ has only a component in $T^*M$, whereas $\sigma_E(\underline X)$ has both $TM$ and $T^*M$ components. However, if we further impose the following \emph{isotropic condition} on $\sigma_E$:
\begin{equation}
\langle\sigma_E(\uX),\sigma_E(\uY)\rangle_E=0
\qquad
\forall \uX,\uY\in\Gamma(TM)\,.
\end{equation}
Then, for any $\un X\in\Gamma(TM)$ and $(\un Y,\{a_i^Y\})\in\Gamma({\cal L}_{{\cal B}_{ij}})$, we have
\begin{equation}
\langle \sigma_E(\uX),(\un Y,\{a_i^Y\})\rangle=\omega_E((\un Y,\{a_i^Y\}))(\un X)\,,
\end{equation}
which means $\omega_E$ can be identified as the pullback $\sigma_E^*$. Then, given an isotropic split, we have the exact sequences going in both directions:
\begin{equation}
\label{LCijSESsplit}
\begin{tikzcd}
0
\arrow{r}
&
\Omega^1(M)
\arrow{r}{\rho_E^*}
\arrow[bend left]{l}
&
{\cal L}_{{\cal B}_{ij}}
\arrow{r}{\rho_E}
\arrow[bend left]{l}{\sigma_E^*}
&
\Gamma(TM)
\arrow{r}
\arrow[bend left]{l}{\sigma_E}
&
0\,.
\arrow[bend left]{l}
\end{tikzcd}
\end{equation}
Given an isotropic split, its curvature can be represented by a closed 3-form $H$
\begin{equation}
R^{\sigma_E}=[\sigma_E(\uX),\sigma_E(\uY)]_E-\sigma_E([\uX,\uY]_{TM})=\rho_E^*(i_{\uY}i_{\uX}H)\,,
\end{equation}
and hence $H$ may be referred to as the curvature of the exact Courant algebroid. The failure of $\sigma_E$ to be a morphism is measured by the curvature $H$. Then, in each local trivialization $E_i\simeq TU_i\oplus T^*U_i$, the bracket on sections is given by the \emph{twisted Courant bracket}:
\begin{equation}
\label{twisted-Courant-local}
[\sigma_E(X)+a^X_i,\sigma_E(Y)+a^Y_i]_{E_i}=\sigma_E([X,Y]_{TM})+\mathcal L_Xa^Y_i-\mathcal L_Ya^X_i-\frac{1}{2}\td(i_Xa^Y_i-i_Ya^X_i)+i_Yi_X H_i\,.
\end{equation}

\paragraph{The \v Severa class.}
The \v Cech 1-cocycle of closed 2-forms $\{{\cal B}_{ij}\}$ determines a cohomology class
\begin{equation}
[{\cal B}_{ij}]\in H^1(M,\Omega^2_{\mathrm{cl}})\,.
\end{equation}
This determines a class in de~Rham cohomology $[H]\in H^3_{\mathrm{dR}}(M)$, known as the \v Severa class of the exact Courant algebroid, which classifies exact Courant algebroids up to isomorphism.\footnote{See \cite{Chen_2013} for a generalization to the non-exact case.} Equivalently, it is the cohomology class of the closed 3-form curvature $H$ obtained after choosing a local isotropic split. This class measures the obstruction to globally trivializing the local Courant algebroid data $TU_i\oplus T^*U_i$ so that the gluing 2-forms ${\cal B}_{ij}$ disappear. 

\subsection{The Exact Courant Algebroid Associated to a Gerbe}
Now suppose we have a $U(1)$-gerbe with connective structure, characterized by the data $(g_{ijk},C_{ij})$. It is obvious that $\td C_{ij}$ is a closed 2‑form, and from $(\delta C)_{ijk} = \td \ln g_{ijk}$ it follows that $\delta \td C=0$. Thus, we can define an exact Courant ${\cal L}_{\td C_{ij}}$ by taking
\begin{align}
{\cal B}_{ij}=\td C_{ij}\,.
\end{align}
This is the exact Courant algebroid naturally derived from a gerbe with connective structure, on which the local 2-forms $B_i$ introduced in the previous section now acquire a second, independent geometric meaning.\footnote{Strictly speaking, the forms $C_{ij}$ and $B_i$ are $\mathfrak u(1)$-valued, so one should write $T^*M\otimes \mathfrak u(1)$ rather than $T^*M$. Since $\mathfrak u(1)$ is a one-dimensional real vector space, we will suppress this distinction.}

\paragraph{Split and 2-form gauge field.}
Suppose we are given a collection of local 2-forms $B=\{B_i\in\Omega^2(U_i,\mathfrak u(1))\}$. We define the horizontal lift of a vector field $\uX$ by specifying $a_i^{X^H}=i_X B_i$, i.e.,
\begin{equation}
\label{Courantsplit}
\sigma_E(\uX)=(\uX,\{i_XB_i\})\,.
\end{equation}
We can immediately see that the gluing relation \eqref{CApatchtrans} requires that
\begin{equation}
B_i-B_j=\td C_{ij}\,.
\end{equation}
Thus, the local 2-form $B_i$ determining a split of the exact Courant algebroid ${\cal L}_{\td C_{ij}}$ is automatically a curving defined in \eqref{deltaB}. Moreover, due to the antisymmetry of $B_i$, we have
\begin{equation}
\langle\sigma_E(\uX),\sigma_E(\uY)\rangle=(i_Xi_YB_i+i_Yi_XB_i)=0\,,
\end{equation}
and hence this split is isotropic. Then, we can write down the curvature 3-form of the 2‑form gauge field $B_i$ as
\begin{align}
\label{gerbecurvature}
H_i\equiv \td B_i\,,
\end{align}
which can be derived from the curvature of $\sigma_E$ or $\omega_E$:
\begin{align}
R^{\sigma_E}(\uX,\uY)=-j_E(R^{\omega_E}((\uX,\{\un a^{X_H}_{i}\}),(\uY,\{\un a^{Y_H}_{i}\})))=(0,\{i_Yi_XH_{i}\})\,.
\end{align}

It is easy to see from $\delta B =\td C$ that $\delta H_i = 0$, namely $H_i = H_j$ on the intersection of any $U_i$ and $U_j$. Hence, the local 3‑forms $H_i$ glue to a global closed 3‑form $H \in \Omega^3(M)$, called the \v Severa form of the exact Courant algebroid $\mathcal L_{\td C_{ij}}$, and its cohomology class $[H]\in H^3(M,\mathbb{R})$ gives the \v Severa class of this Courant algebroid. On the other hand, we also recognize $H \in \Omega^3(M)$ as the curvature of the $U(1)$‑gerbe with connective structure, representing the Dixmier--Douady class in $H^3(M,\mathbb{Z})$. Thus, in this case the \v Severa class is \emph{integral} in the sense that it comes from an integral cohomology class, which need not be true for a general exact Courant algebroid $\mathcal L_{{\cal B}_{ij}}$ \cite{gualtieri2011generalized}. In other words, exact Courant algebroids associated with $U(1)$‑gerbes with connective structure are precisely those whose \v Severa class lies in the image of $H^3(M,\mathbb{Z})\to H^3(M,\mathbb{R})$.

At this point the geometric role of all local data has been identified. The gerbe cocycle $c_{ijk}=\ln g_{ijk}$ defines the highest \v Cech component, the connective structure $C_{ij}$ defines the split of the Lie 2-algebroid ${\cal L}_{g_{ijk}}$, and the local 2-forms $B_i$ define the isotropic split of the associated exact Courant algebroid ${\cal L}_{\td C_{ij}}$. We now collect these ingredients and formulate the higher Russian formula.

\section{Higher Russian Formula and the BV--BRST Complex}
\label{sec:Russian}
Given a $U(1)$-gerbe with connective structure and curving, a choice of local trivializations determines local \v Cech--de~Rham data
\begin{equation}
(B_i,C_{ij},c_{ijk})\,,
\end{equation}
where
\begin{equation}
B_i\in C^0({\cal U},\Omega^2(M))\,,\qquad
C_{ij}\in C^1({\cal U},\Omega^1(M))\,,\qquad
c_{ijk}\in C^2({\cal U},\Omega^0(M))\,.
\end{equation}
These local forms arise from the two geometric structures introduced in the previous sections. The connective structure $C_{ij}$ induces a split of the Lie 2-algebroid ${\cal L}_{g_{ijk}}$, while the curving $B_i$ determines an isotropic split of the associated exact Courant algebroid ${\cal L}_{\td C_{ij}}$.

We have seen that from the split Lie 2-algebroid one obtains the descent relation
\begin{equation}
\label{ghostdescentidentity}
(\td+\delta)(C-c)=\delta B\,,
\end{equation}
while from the split exact Courant algebroid one obtains the curvature identity
\begin{equation}
\label{curvingidentity}
\td B=H\,,
\end{equation}
where the local 3-forms $\td B_i$ glue to a globally defined closed 3-form curvature. Combining \eqref{ghostdescentidentity} and \eqref{curvingidentity}, we obtain the \emph{higher Russian formula}
\begin{equation}
\label{HigherRussianFormula}
\boxed{(\td+\delta)(B-C+c)=H}\,.
\end{equation}
It is convenient to introduce the total field
\begin{equation}
\omega_{\rm tot}\equiv B-C+c
\end{equation}
in the total \v Cech--de~Rham complex. Then \eqref{HigherRussianFormula} can be written as
\begin{equation}
(\td+\delta)\omega_{\rm tot}=H\,.
\end{equation}
This is the direct higher analogue of the ordinary Russian formula \eqref{Russianf1}, with $\omega=b-\varpi$ replaced by the total field $\omega_{\rm tot}$.

Expanding \eqref{HigherRussianFormula} by \v Cech degree recovers the local relations
\begin{equation}
(\delta c)_{ijkl}=0\,,\qquad
(\delta C)_{ijk}=(\td c)_{ijk}\,,\qquad
(\delta B)_{ij}=(\td C)_{ij}\,,\qquad
\td B_i=H_i\,.
\end{equation}
Thus, the higher Russian formula packages the full hierarchy of local descent relations together with the global curvature into a single identity. From the viewpoint of physics, the first three identities are recognized as the familiar BRST transformation laws of an Abelian reducible gauge system, or equivalently as the minimal BV--BRST differential on the field-ghost tower. In particular, they identify $C$ as the 1-form ghost associated to the 2-form gauge field $B$, and $c$ as the corresponding ghost-for-ghost. The total field $\omega_{\rm tot}=B-C+c$ may therefore be viewed as the extended field $\hat B$ encoding the full field-ghost hierarchy (see Appendix \ref{App:BV}), while the final identity $\td B_i=H_i$ expresses the globally defined 3-form curvature.

This also clarifies the relation between the \v Cech--de~Rham bicomplex in the gerbe case and the BV--BRST complex. Recall that in the 0-form symmetry case, the \v Cech--de~Rham bicomplex provides a local model for the BRST complex of an irreducible gauge theory, where the \v Cech differential $\delta$ plays the role of the BRST differential, and the degree-1 \v Cech cocycle corresponds to the ghost field. In the 1-form symmetry case, the gauge symmetry is reducible, and the corresponding local complex naturally represents the minimal field-ghost sector of the BV--BRST complex. The \v Cech degree records the ghost hierarchy: $C$ is the 1-form ghost and $c$ is the 0-form ghost-for-ghost. In this sense, the role of the \v Cech--de~Rham bicomplex in the gerbe case is the precise higher analogue of its role in the ordinary principal bundle case. For a principal bundle it models the BRST complex associated with the Atiyah Lie algebroid; for gerbes it models the corresponding higher field-ghost BV--BRST complex associated with the Lie 2-algebroid, while the curving simultaneously determines the split of the associated exact Courant algebroid encoding the 3-form curvature. 

\section{Anomaly Polynomials for 1-Form Symmetries}
\label{sec:descent-higherform}
\subsection{Descent Equations and Anomaly Polynomials}
Building on \cite{Jia:2023tki}, the BRST complex for an ordinary gauge symmetry can be formulated geometrically using the Atiyah Lie algebroid of a principal bundle. In a local trivialization, the algebroid Chern--Weil construction produces characteristic classes and their Chern--Simons transgression forms, whose bi-degree expansion yields the descent equations \cite{Stora1977}, and hence the Wess--Zumino consistency condition \cite{wess1971consequences} for the consistent anomaly. In the present work we have derived the higher Russian formula and the corresponding BV--BRST bicomplex for a $U(1)$ gerbe in \v Cech--de~Rham data. We now explain how the same descent mechanism produces anomaly polynomials for 1-form symmetries with 2-form gauge fields.

\paragraph{Anomaly descent for ordinary symmetries}
For an ordinary gauge symmetry with local gauge field $b$ and curvature $F=\td b+\frac{1}{2}[b,b]_L$, a characteristic class is a closed form constructed from the curvature
\begin{equation}
I_{d+2}(F)={\cal Q}(F,\ldots,F)\in\Omega^{d+2}(M)\,,
\end{equation}
where $\cal Q$ is a symmetric invariant polynomial valued in the Lie algebra $\mathfrak{g}$ of the symmetry. Locally, one can write
\begin{equation}
I_{d+2}(F)=\td\mathrm{CS}_{d+1}(b)\,,
\end{equation}
where $\mathrm{CS}_{d+1}(b)$ is the corresponding Chern--Simons transgression form.

In the BRST complex coming from the trivialized Atiyah Lie algebroid, the algebroid differential $\hat\td$ decomposes as $\td+\ts$,
and the connection reform $\omega=b-\varpi$ realizes the extended connection, whose total curvature coincides with $F$ by the Russian formula. Expanding $\mathrm{CS}_{d+1}(\omega)$ by the de~Rham form degree $p$ and the ghost number $q$ yields a sequence of descent equations
\begin{equation}
\td \alpha^{(p,q)}+\ts\alpha^{(p+1,q-1)}=0\,,\qquad p+q=d+1\,,
\end{equation}
and in particular the Wess--Zumino consistency condition
\begin{equation}
\td \alpha^{(d-1,2)}+\ts\alpha^{(d,1)}=0\,.
\end{equation}
The consistent anomaly of a quantum field theory on a $d$-dimensional spacetime is then encoded by $\alpha^{(d,1)}$, with $\alpha^{(d,1)}\neq\td \alpha^{(d-1,1)}+\ts\alpha^{(d,0)}$. In other words, the anomaly lives in $H^{d,1}(\td|\ts)$, the ghost number one sector of the BRST cohomology. $I_{d+2}$ provides a topological invariant called the \emph{anomaly polynomial}, while $\alpha^{(d,1)}$ is the consistent anomaly descending from it.

\paragraph{Anomaly descent for 1-form symmetries}

A key point, already familiar from the Atiyah Lie algebroid description of the BRST complex, is that the Chern--Weil construction depends only on the existence of a closed differential subalgebra generated by a ``connection'' and its ``curvature,'' rather than on any particular formulation of the total space of a principal bundle. In the present higher-form setting, the higher Russian formula \eqref{HigherRussianFormula} exhibits the 3-form $H$ as a $\td+\delta$-closed, ghost-free curvature associated to the total BV--BRST complex. Consequently, invariant polynomials in $H$ define cocycles in the total bicomplex, and admit local Chern--Simons transgression representatives exactly as in ordinary Chern--Weil theory. In particular, the cohomological characterization of anomalies via descent continues to apply in this higher algebroid formulation.

At the purely local/perturbative level, the analogue of the anomaly polynomial is now a closed gauge-invariant $(d+2)$-form constructed from $H$ (and possibly gravitational data from the frame bundle or its associated tangent bundle). In the simplest case of a pure 1-form symmetry background, this means a polynomial of the schematic form
\begin{equation}
I_{d+2}(H)\in\Omega^{d+2}(M)\,,\qquad \td I_{d+2}(H)=0\,.
\end{equation}
One can also consider cases involving Pontryagin classes of $TM$ as polynomials of the spacetime curvature 2-form $R$:
\begin{equation}
I_{d+2}(H,R)\in\Omega^{d+2}(M)\,,\qquad \td I_{d+2}(H,R)=0\,,
\end{equation}
which give rise to gauge-gravitational mixed anomalies.

Locally, $I_{d+2}(H)$ admits a Chern--Simons transgression $(d+1)$-form $\mathrm{CS}_{d+1}$ built from the BV--BRST data $(B,C,c)$. Expanding $\mathrm{CS}_{d+1}$ by the de~Rham form degree $p$ and \v Cech degree (ghost number) $q$ yields the descent equations
\begin{equation} \label{eq:descenteq1form}
\td \alpha^{(p,q)}+\delta\alpha^{(p+1,q-1)}=0\,,\qquad p+q=d+1\,,
\end{equation}
including in particular the Wess--Zumino consistency condition
\begin{equation}
\td \alpha^{(d-1,2)}+\delta\alpha^{(d,1)}=0\,.
\end{equation}
This naturally makes $\alpha^{(d,1)}$ with $\alpha^{(d,1)}\neq\td\alpha^{(d-1,1)}+\delta\alpha^{(d,0)}$ a candidate for the consistent anomaly for a 1-form symmetry in a theory on a $d$-dimensional spacetime. Therefore, the anomaly of 1-form symmetry descending from $I_{d+2}$ lives in $H^{d,1}(\td|\delta)$, the ghost number one sector of the BV--BRST cohomology.

Since $H$ is a form of odd degree, we have $H\wedge H = 0$ in the Abelian case.\footnote{This statement is only at the level of de~Rham anomaly polynomials. At the level of integral cohomology, the cup square $\iota\smile\iota\in H^6(K(\mathbb Z,3),\mathbb Z)$ is a nonzero $\mathbb Z_2$-torsion class. Such torsion classes would potentially correspond to global anomalies, which are invisible to the local descent formalism. However, according to bordism computations, $\iota\smile\iota$ itself does not contribute to a global anomaly. A detailed analysis of these issues will be given in \cite{PartII}.} Thus, a single Abelian 1-form symmetry does not admit a perturbative self anomaly captured by a characteristic class built purely from $H$ at the level of local anomaly polynomials. Nontrivial anomaly polynomials can only arise from mixed structures, for instance for two $U(1)$ 1-form symmetries with curvatures $H_1,H_2$ one may consider $I_6 \sim H_1\wedge H_2$, which leads to a mixed anomaly under the corresponding higher-form gauge transformations. In the following, we will provide concrete physical examples of these anomalies.

The descent formalism produces the perturbative (local) anomaly conveniently once the anomaly polynomial is specified. To place the anomaly polynomials in a more general framework, in a companion paper \cite{PartII} we apply the bordism analysis as the classification framework for invertible phases with background $B$-field. This will also allow us to detect the non-perturbative (global) anomaly. 

\subsection{Examples}
\subsubsection{Maxwell Theory in $4d$}
We first consider the 4-dimensional pure $U(1)$ Maxwell theory with the action~\cite{Gaiotto:2014kfa,Luo:2023ive}
\be
\label{Max-free-action}
S_{\rm Max}=\int_{M_4} \frac{1}{2e^2}F\wedge *F\,.
\ee
The theory has an electric $U(1)_E^{(1)}$ 1-form symmetry with the current
\be
\label{Max-e-current}
j_E=\frac{1}{e^2}F
\ee
and a magnetic $U(1)_M^{(1)}$ 1-form symmetry with the current
\be
\label{Max-m-current}
j_M=\frac{1}{2\pi}*F\,.
\ee
Note that the action of $U(1)_E^{(1)}$ can also be thought of as the shifting
\be
A\rightarrow A+\lambda_E
\ee
with a flat connection $\lambda_E$. Note that the action \eqref{Max-free-action} does not have an ABJ anomaly \cite{adler1969axial,bell1969pcac} for $U(1)_E^{(1)}\times U(1)_M^{(1)}$.

It is known that the $U(1)$ Maxwell theory suffers from a mixed 't Hooft anomaly which obstructs the gauging of the whole 1-form global symmetry $U(1)_E^{(1)}\times U(1)_M^{(1)}$, due to the fact that the gauged action
\be
S_{\rm gauged}=\int_{M_4} \frac{1}{2e^2}(F+B_E)\wedge *(F+B_E)+\int_{M_4}\frac{1}{2\pi}(F+B_E)\wedge B_M
\ee
is not gauge invariant under the gauge transformation
\be
\delta_{\lambda_E}A=\lambda_E\,,\qquad \delta_{\lambda_E}B_E=-\td\lambda_E\,,\qquad \delta_{\lambda_M}B_M=-\td\lambda_M\,.
\ee
The failure of gauge invariance reads
\be
\delta S_{\rm gauged}=-\int_{M_4}\frac{1}{2\pi}B_E\wedge \td\lambda_M=\frac{1}{2\pi}\int_{M_4}H_E\wedge \lambda_M \,,
\ee
where in the second equality we assumed that $M_4$ is closed. This anomaly is encoded by the gauge-invariant 't Hooft anomaly polynomial 6-form\footnote{A careful analysis at the level of differential cohomology can be found in~\cite{Hsieh:2020jpj}, where the anomaly result is the same and subtleties regarding background gauge field couplings are clarified.}
\be
\label{Maxwell-I6}
I_6=\frac{1}{(2\pi)^2}\td B_E\wedge \td B_M=\frac{1}{(2\pi)^2}H_E\wedge H_M\,.
\ee
A Chern--Simons transgression form satisfying $\td {\rm CS}_5=I_6$ is
\be
{\rm CS}_5=-\frac{1}{(2\pi)^2}H_E\wedge B_M\,.
\ee
Indeed, under the gauge variation $\delta_{\lambda_M}B_M=-\td\lambda_M$ we have
\be
\delta_{\lambda_M}{\rm CS}_5
=
\frac{1}{(2\pi)^2}H_E\wedge \td\lambda_M
=
-\td\left(\frac{1}{(2\pi)^2}H_E\wedge \lambda_M\right)\,,
\ee
which on the boundary gives the mixed anomaly in four dimensions. Equivalently, one may use the representative
\be
{\rm CS}'_5=\frac{1}{(2\pi)^2}B_E\wedge H_M\,,
\ee
which differs from ${\rm CS}_5$ by a boundary local counterterm and places the anomaly in the $U(1)_E^{(1)}$ gauge variation.

Interestingly, this 't Hooft anomaly also admits an alternative interpretation as an ABJ anomaly. Let us consider the action of \eqref{Max-free-action} after gauging the $U(1)_E^{(1)}$ 1-form symmetry, with the action
\be
S_{\text{2-group}}=\int_{M_4} \frac{1}{2e^2}(F+B_E)\wedge *(F+B_E)\,.
\ee
Note that such an action can be regarded as the action for a strict Lie 2-group gauge theory with $G=H=U(1)$, the group homomorphism $t:H\rightarrow G$ as an identity map, and the trivial group homomorphism $\alpha:G\rightarrow \mathrm{Aut}(H)$ ~\cite{Baez:2010ya,Kapustin:2013uxa}. The gauge transformation rules are exactly
\be
\delta_{\lambda_E}A=\lambda_E\,,\qquad \delta_{\lambda_E}B_E=-\td\lambda_E\,.
\ee
For this theory, the current \eqref{Max-m-current} is no longer conserved, as
\be
\td*j_M=\frac{1}{2\pi}\td B_E=\frac{1}{2\pi}H_E
\ee
is non-vanishing after $U(1)_E^{(1)}$ is gauged. The non-conservation $\td*j_M$ again signifies the presence of an ABJ anomaly for the $U(1)_M^{(1)}$ 1-form global symmetry. The anomaly polynomial \eqref{Maxwell-I6} represents the degree-6 class in $H^6(B^2U(1)_E\times B^2U(1)_M,\mathbb Z)$ given by the product of the two Dixmier--Douady classes.

\subsubsection{Maxwell Theory in $5d$}

We next consider the 5-dimensional pure $U(1)$ Maxwell theory with action
\begin{equation}
S_{\rm Max}=\int_{M_5} \frac{1}{2e^2}F\wedge *F\,.
\end{equation}
As in four dimensions, this theory has an electric $U(1)^{(1)}$ 1-form symmetry whose charged objects are Wilson lines. We assume that there are no dynamical electrically charged particles, so that this 1-form symmetry is preserved. The symmetry acts on the gauge field by the shift
\begin{equation}
A \to A+\lambda^{\mathrm{flat}}\,,
\end{equation}
where $\lambda^{\mathrm{flat}}$ is a flat connection satisfying $\td \lambda^{\mathrm{flat}}=0$.

In five dimensions, a natural anomaly involving this symmetry is a mixed gauge-gravitational anomaly with anomaly polynomial
\begin{equation}
I_7=H\wedge X_4\,,
\end{equation}
where $H$ is the curvature of the background $U(1)$-gerbe field $B$ and $X_4$ is a closed gravitational characteristic form, for example $X_4=p_1(TM)$.\footnote{On spin manifolds, $\frac{1}{2} p_1(TM)$ is an integral class, so one may consider $X_4=\frac{1}{2} p_1(TM)$.}
Since this anomaly polynomial factorizes, it admits a Green--Schwarz mechanism interpretation.

To gauge the electric 1-form symmetry, one introduces a background 2-form gauge field $B$ with curvature $H=\td B$. Then, the ordinary curvature $F=\td A$ can be replaced by the fake curvature
\begin{equation}
\tilde F=F+B\,,
\end{equation}
which satisfies the modified Bianchi identity
\begin{equation}
\td \tilde F=H\,.
\end{equation}
The fake curvature is invariant under the 1-form gauge transformation
\begin{equation}
\delta A=\lambda\,,\qquad \delta B=-\td\lambda\,.
\end{equation}
for an arbitrary connection $\lambda$.

The factorized form of $I_7$ implies that one may locally choose a Chern--Simons transgression $6$-form
\begin{equation}
\mathrm{CS}_6=B\wedge X_4\,,
\end{equation}
satisfying
\begin{equation}
\td \mathrm{CS}_6=\td(B\wedge X_4)=H\wedge X_4=I_7\,.
\end{equation}
Under the $1$-form gauge transformation, this transgression form varies as
\begin{equation}
\delta \mathrm{CS}_6=-\td(\lambda\wedge X_4)\,,
\end{equation}
and therefore induces the boundary variation
\begin{equation}
\delta S=-\int_{M_5}\lambda\wedge X_4\,.
\end{equation}
This is precisely the descent variation associated with the anomaly polynomial $I_7$.

Equivalently, the same structure may be represented through a Green--Schwarz type topological coupling on a closed $5$-manifold $M_5$:
\begin{equation}
S_{\mathrm{GS}}=-\int_{M_5}A\wedge X_4
=-\int_{M_5}F\wedge \mathrm{CS}^{\rm grav}_3\,,
\end{equation}
where $\mathrm{CS}^{\rm grav}_3$ is a local gravitational Chern--Simons form satisfying locally
\begin{equation}
\td \mathrm{CS}^{\rm grav}_3 = X_4\,.
\end{equation}
The non-invariance of $S_{\mathrm{GS}}$ under the 1-form shift of $A$ reproduces the same anomaly by descent. If there exists a $6$-manifold $M_6$ with $\partial M_6=M_5$, one may equivalently write this as a bulk Wess--Zumino term
\begin{equation}\label{eq:6dWZ}
S_{\mathrm{GS}}=-\int_{M_6} F\wedge X_4\,,
\end{equation}
which avoids the caveat that $A_1$ or $\mathrm{CS}^{\rm grav}_3$ is defined only locally.

We conclude by noting that this anomaly has further physical implications. On the one hand, the coupling $S_{\mathrm{GS}}$ implies that magnetic string defects in five dimensions carry additional topological data beyond their magnetic charge. On the other hand, this anomaly admits a natural top-down realization in string theory through dimensional reduction of higher dimensional couplings. These directions will be developed in detail in the companion paper \cite{PartII}.

\section{Conclusions}
\label{sec:conclusions}
In this work, we studied the geometry underlying the BV--BRST description of a $U(1)$ 1-form gauge symmetry. We began by reformulating the Atiyah Lie algebroid, which underlies the BRST complex of ordinary gauge symmetry, in terms of its local \v Cech presentation ${\cal L}_{g_{ij}}$. This perspective is particularly useful because, for a gerbe, the higher analogue of the Atiyah Lie algebroid is no longer described by an ordinary vector bundle. The natural higher symmetry object therefore appears directly in the \v Cech description.

Since the background field for a $U(1)$ 1-form symmetry comes from a gerbe, which is naturally formulated in \v Cech data, we developed the corresponding higher structure directly in the \v Cech picture. Starting from the gerbe cocycle, we introduced the local infinitesimal symmetry object ${\cal L}_{g_{ijk}}$, which should be regarded as a \v Cech presentation of the Lie 2-algebroid associated to the gerbe. From the physical viewpoint, we identify the connective data $C$ with the 1-form ghost, the \v Cech 2-cocycle $c$ with the ghost-for-ghost, and the curving $B$ with the 2-form gauge field. Given a gerbe with connective structure, one can further construct an associated exact Courant algebroid, where the same gauge field $B$ arises from a choice of split.

Using these ingredients, we derived the higher Russian formula \eqref{HigherRussianFormula}, which packages the local gerbe data and the global 3-form curvature into a single identity. Its expansion by \v Cech degree reproduces the hierarchy of local descent relations and identifies the resulting \v Cech--de~Rham bicomplex with the minimal BV--BRST field-ghost complex of an Abelian reducible gauge system. We then showed that the curvature $H$ plays the role of the ghost-free total curvature in the higher setting, allowing one to construct anomaly polynomials and derive the corresponding consistent anomaly by descent. Finally, we illustrated the formalism in two explicit examples: the mixed anomaly of electric and magnetic $U(1)$ 1-form symmetries in $4d$ Maxwell theory, which can be viewed as an ABJ anomaly, and the mixed gauge-gravitational anomaly in $5d$ Maxwell theory, which admits a Green--Schwarz type interpretation.

There are several directions in which the present construction could be developed further. First, from a geometric viewpoint, our treatment of the Lie 2-algebroid is given in a local \v Cech presentation rather than in a direct analogue of the vector bundle description of the Atiyah Lie algebroid. It would be desirable to identify the precise global geometric object underlying this presentation, possibly in relation to higher principal bundles and their infinitesimal symmetries \cite{JARDINE_LUO_2006,Ginot:2008fia,Nikolaus_2014a,Nikolaus_2014b,sheng2017string,waldorf2017,Zucchini:2019pbv}. Closely related to this is the fact that, in our derivation of the higher Russian formula, both the Lie 2-algebroid and the associated Courant algebroid play essential roles. An important open question is therefore whether there exists a single geometric structure that unifies the two and from which the higher Russian formula follows directly. If so, it would be natural to ask whether it carries an intrinsic cohomology theory, analogous to the cohomology of the Atiyah Lie algebroid defined by $\hat\td$, and whether the total field $\omega_{\rm tot}$ can be interpreted as a geometric object on it, just as the connection form $\omega$ is naturally associated with the Atiyah Lie algebroid.

The recent development in \cite{Bunk:2026bha} provides a natural framework in which these questions may be formulated globally. In that work, the Atiyah $L_\infty$-algebroid of a principal $\infty$-bundle is introduced, $p$-form connections are interpreted as order-$p$ splits, the \v Cech--Deligne description of connections on higher $U(1)$-bundles is recovered, and derived higher symmetries of higher $U(1)$-bundles are related to higher Courant algebroids. From the perspective of the present paper, this suggests that our local \v Cech Lie 2-algebroid should arise as a presentation of a global Atiyah $L_\infty$-algebroid of a gerbe, while the associated Courant algebroid may encode the additional curvature data entering the higher Russian formula. The present work emphasizes a complementary physical aspect of these structures: the emergence of the BV--BRST field-ghost tower, the higher Russian formula, and anomaly descent for $U(1)$ 1-form symmetries.

More generally, the present construction should admit a direct higher-form generalization. For a pure $p$-form gauge symmetry, our local \v Cech construction should naturally extend to a corresponding higher algebroid structure associated with a higher gerbe. However, our construction also depends crucially on the associated exact Courant algebroid, and so a systematic generalization would require the appropriate higher analogue of this structure as well. The framework of \cite{Bunk:2026bha}, especially its treatment of $p$-form connections on principal $\infty$-bundles and its relation between higher $U(1)$-bundles and higher Courant algebroids, suggests a promising route toward such a generalization.

Furthermore, the descent data constructed in this paper encode the consistent anomaly, since in the case of ordinary gauge symmetry, the \v Cech--de~Rham complex arises from the consistent splitting of the Atiyah Lie algebroid. It has been shown in \cite{Jia:2023tki} that one may also choose a covariant splitting of the Atiyah Lie algebroid, whose corresponding cohomological description leads directly to the covariant anomaly \cite{Bardeen:1984pm}. It would be interesting to understand whether an analogous covariant splitting exists in the Lie 2-algebroid setting and, if so, how the corresponding covariant anomaly is represented.

There are also a number of physical extensions. One natural direction is to couple the gerbe sector to an ordinary Yang--Mills theory, where the interplay between 1-form and 0-form symmetries may be organized into a higher-group structure \cite{Gukov:2013zka,Kapustin:2013uxa,Cordova:2018cvg,Delcamp:2018wlb,Benini:2018reh,Tanizaki:2019rbk,Iqbal:2020lrt,Cordova:2020tij,DelZotto:2020sop}. Another is to incorporate matter fields. In the higher-form setting, this should naturally include string-like charged objects, which may be regarded as a kind of 2-matter and may require an appropriate notion of higher representation for the underlying higher algebroid structure. 

\section*{Acknowledgements}
We would like to thank Hank Chen, Jingyuan Chen, Zhuo Chen, Zheng-Cheng Gu, Marc Klinger, Tian Lan, Honglei Lang, Saito Shota, Yuji Tachikawa, Qing-Rui Wang, Piljin Yi, Hao Y.~Zhang for discussions. WJ is supported by the Research Grants Council of Hong Kong Special Administrative Region of China (Project No.~14302725) under the scheme of General Research Fund. YNW is supported by National Natural Science Foundation of China under Grant No.~12422503. YZ is supported by WPI Initiative, MEXT, Japan at Kavli IPMU, the University of Tokyo.

\appendix

\section{Atiyah and Courant Algebroids}
\label{App:ALACourant}
\subsection{Atiyah Lie Algebroids}
\label{App:ALA}
A \emph{Lie algebroid} is a vector bundle $A$ over $M$, along with a bundle map $\rho: A \rightarrow TM$ called the anchor map, and a bracket $[\cdot,\cdot]_A$ satisfying
\begin{enumerate}[(a)]
\item $[[\umX,\umY]_A, \umZ]_A + [[\umZ, \umX]_A, \umY]_A + [[\umY, \umZ]_A, \umX]_A = 0. \qquad\forall\umX,\umY,\umZ\in\Gamma(A)$;
\par
\item $[f\umX,g\umY]_A=fg[\umX,\umY]_A+f(\rho(\umX)g)\umY-g(\rho(\umY)f)\umX\,.\qquad\forall\umX,\umY\in\Gamma(A),\quad f\in C^{\infty}(M)$.
\end{enumerate}
The first condition implies that the bracket $[\cdot,\cdot]_A$ satisfies the \emph{Jacobi identity}, and the second implies that the anchor is compatible with the bracket. Conditions $(a)$ and $(b)$ together imply that $\rho$ is a morphism of brackets between $A$ and the tangent bundle $TM$. We can equivalently state this as the fact that the curvature of the map $\rho$ vanishes:
\begin{equation}
R^\rho(\umX,\umY)\equiv[\rho(\umX),\rho(\umY)]_{TM}-\rho([\umX,\umY]_A)=0\,.
\end{equation}

If $\rho$ is surjective, then $A$ is called a \emph{transitive Lie algebroid}. In this case, we have the short exact sequence:
\begin{equation} \label{Transitive Lie Algebroid}
\begin{tikzcd}
0
\arrow{r} 
& 
L
\arrow{r}{j} 
&
A
\arrow{r}{\rho} 
& 
TM
\arrow{r} 
&
0 \,,
\end{tikzcd}
\end{equation}
where $j$ is an inclusion of a vector bundle $L$ over $M$ into $A$, whose image is the kernel of $\rho$. $L$ is called the isotropy bundle of $A$, and its image $\text{im}(j) = \text{ker}(\rho)\subset A$ is referred to as the vertical sub-bundle $V\subset A$. It is natural to require that $j$ is a morphism, i.e.\
\begin{equation}
  R^j(\umu,\unu)\equiv  [j(\underline{\mu}), j(\underline{\nu})]_A-j([\underline{\mu}, \underline{\nu}]_L) = 0\,,\qquad\forall\umu,\unu\in\Gamma(L)\,,
\end{equation}
Hence, we may consider $V$ as a ``copy'' of the bundle $L$ embedded in $A$. Note that the bracket $[\cdot,\cdot]_L$ on $L$ induced by $[\cdot,\cdot]_A$ is linear, and thus $L$ defines a ``bundle of Lie algebras'' over $M$ as each fiber at $p\in M$ is isomorphic to a single Lie algebra, called the isotropy bundle of $A$. 
\par

The canonical example of a transitive Lie algebroid is the \emph{Atiyah Lie algebroid}. Given a principal $G$-bundle $P(M,G)$, one immediately obtains an associated Atiyah Lie Algebroid defined by the short exact sequence:
\begin{equation} \label{Atiyah Lie Algebroid}
\begin{tikzcd}
0
\arrow{r} 
& 
L = P\times_{\text{Ad}} \mathfrak{g}
\arrow{r}{j} 
&
A = TP/G
\arrow{r}{\rho \;=\; \pi_*} 
& 
TM
\arrow{r} 
&
0\,.
\end{tikzcd}
\end{equation}
The isotropy algebra of an Atiyah Lie algebroid is the adjoint bundle of Lie algebras, i.e.\ $L=P\times_{\text{Ad}} \mathfrak{g}$, whose elements are precisely the local gauge transformations. 

\paragraph{Connection and curvature on $A$.} As is the case with a principal bundle, a transitive Lie algebroid is endowed with a vertical sub-bundle $V\subset A$, but does not possess a canonically defined horizontal sub-bundle $H\subset A$ such that $A=H\oplus V$ globally. Each choice of a horizontal sub-bundle $H$ of $A$ defines a connection on $A$, which can be characterized by two bundle maps $\omega: A \rightarrow L$ and $\sigma: TM \rightarrow A$. $\omega:A\to L$ is called a \emph{connection reform} and satisfies
\begin{equation}
\omega(j(\umu))=-\umu\,.\qquad \forall\umu\in\Gamma(L)\,.
\end{equation}
Once an $\omega$ is chosen, its kernel defines a horizontal sub-bundle $H\subset A$. On the other hand, one can define a \emph{connection} (or a \emph{split}) $\sigma:TM\to A$ satisfying
\begin{equation}
\rho(\sigma(\uX))=\uX\,,\qquad\forall\uX\in TM\,,
\end{equation}
The image of $\sigma$ defines the horizontal sub-bundle $H$, meaning each $\omega$ is paired up with a unique $\sigma$ by the following relation:
\begin{equation}
\text{ker}(\omega)=\text{im}(\sigma)=H\subset A\,.
\end{equation}
We can see that $\sigma$ and $\omega$ play the same role as the horizontal lift and Ehresmann connection defined on a principal bundle, respectively. It is easy to see that $\omega \circ \sigma = 0$, and so the exact sequence of the algebroid can now go in either direction:
\begin{equation}
\begin{tikzcd}
0
\arrow{r} 
& 
L
\arrow{r}{j} 
\arrow[bend left]{l} 
& 
A
\arrow{r}{\rho} 
\arrow[bend left]{l}{\omega}
& 
TM
\arrow{r} 
\arrow[bend left]{l}{\sigma}
&
0\,.
\arrow[bend left]{l} 
\end{tikzcd}
\end{equation}

Any transitive Lie algebroid possesses an exterior algebra: 
\begin{equation}
\Omega^\bullet(A)=\bigoplus_{n=0}^{\text{rank}A}\Omega^n(A)\,,
\end{equation}
where $\Omega^{n}(A)\equiv\Gamma(\wedge^nA^*)$ consists of $n$-multi-linear, totally anti-symmetric maps from $A^{\otimes n}$ to $\mathbb{R}$.
The coboundary operator of this exterior algebra, $\hat \td:\Omega^n(A)\to\Omega^{n+1}(A)$, is defined using the Koszul formula: 
\begin{multline} \label{dhat for trivial bundle}
    \hat{\td}\eta(\underline{\mathfrak{X}}_1, \cdots, \underline{\mathfrak{X}}_{p+1}) = \sum_{i} (-1)^{i+1} \rho(\underline{\mathfrak{X}}_i) \eta(\underline{\mathfrak{X}}_1, \cdots, \widehat{\underline{\mathfrak{X}}_i}, \cdots, \underline{\mathfrak{X}}_{p+1}) \\
    + \sum_{i < j} (-1)^{i + j} \eta([\underline{\mathfrak{X}}_i, \underline{\mathfrak{X}}_j]_A, \underline{\mathfrak{X}}_1, \cdots, \widehat{\underline{\mathfrak{X}}_i}, \cdots, \widehat{\underline{\mathfrak{X}}_j}, \cdots, \underline{\mathfrak{X}}_{p+1}) \,,
\end{multline}
where $\eta\in\Omega^{p}(A^*)$, and the hat stands for omission. The exterior algebra $\Omega^{\bullet}(A)$ can be extended to $\Omega^{\bullet}(A,E)$, namely the exterior algebra on $A$ with values in the vector bundle $E$, by introducing a suitable differentiation of sections of $E$. Such a notion comes in the form of a \emph{Lie algebroid representation}, which is a morphism $\phi_{E}: A \rightarrow \text{Der}(E)$ compatible with the anchor. The morphism condition simply means that $\phi_{E}$ has a vanishing curvature:
\begin{equation} \label{Phi E Morphism}
	R^{\phi_{E}}(\un\mX,\un\mY) = [\phi_{E}(\un\mX),\phi_{E}(\un\mY)]_{\text{Der}(E)} - \phi_{E}([\un\mX,\un\mY]_A) = 0\,,\qquad\forall  \un\mX,\un\mY \in A\,.
\end{equation}
The compatibility condition ensures that $\phi_{E}$ maps into a derivation by enforcing the Leibniz-like identity
\begin{equation}
	\phi_{E}(\un\mX)(f \un\psi) = f \phi_{E}(\un\mX)(\un\psi) + \rho(\un\mX)(f) \un\psi\,,\qquad \forall  \un\mX \in A\,,\quad f \in C^{\infty}(M)\,,\quad \un\psi \in E\,.
\end{equation}
Given such a representation, one can modify \eqref{dhat for trivial bundle} into the Koszul formula valid for $\Omega(A;E)$ simply by replacing $\rho$ with $\phi_{E}$:
\begin{align} \label{dhat on E}
    \hat{\td}^E\eta(\un{\mX}_1, \cdots, \un{\mX}_{p+1}) ={}& \sum_{i} (-1)^{i+1} \phi_{E}(\un{\mX}_i) \eta(\un{\mX}_1, \cdots, \widehat{\un{\mX}_i}, \cdots, \un{\mX}_{p+1}) \nn\\
    &+ \sum_{i < j} (-1)^{i + j} \eta([\un{\mX}_i, \un{\mX}_j]_A, \un{\mX}_1, \cdots, \widehat{\un{\mX}_i}, \cdots, \widehat{\un{\mX}_j}, \cdots, \un{\mX}_{p+1}) \,.
\end{align}
The operator $\hat{\td}^E$ can immediately be seen to be nilpotent as a combination of \eqref{Phi E Morphism} and the fact that the bracket on $A$ satisfies the Jacobi identity. For simplicity, later on we will refer to the coboundary operator as $\hat{\td}$ without specifying the representation $E$.

Since the connection reform $\omega$ is an $L$-valued 1-form on $A$, we can now compute its corresponding \emph{curvature reform}, which is an $L$-valued 2-form defined as
\begin{equation}
\label{curvatureOmega}
    \Omega = \hat{\td} \omega + \frac{1}{2}[\omega \wedge \omega]_L\,.
\end{equation}
It can be shown that $\Omega$ contracted with any vertical vectors in $A$ vanishes, i.e.,
\begin{align} \label{CurvatureHorizontal}
\Omega(\umX^H, \umY_V) = \Omega(\umX_V, \umY^H) = \Omega(\umX_V, \umY_V) = 0\,.
\end{align}
This indicates that the curvature reform of a transitive Lie Algebroid is horizontal. This object corresponds to the curvature we defined in \eqref{curvatureeq} in the \v Cech language, whose horizontality as a geometric fact on the algebroid gives rise to the Russian formula in the BRST formalism.

The curvature of a connection on the Lie algebroid $A$ can also be quantified by the curvature of the maps $\omega$ and $\sigma$:
\begin{align}
\label{Rsigma}
    R^{\sigma}(\underline{X}, \underline{Y}) &= [\sigma(\underline{X}), \sigma(\underline{Y})]_A - \sigma([\underline{X}, \underline{Y}]_{TM})\,,\\
    \label{Romega}
    R^{-\omega}(\underline{\mathfrak{X}}, \underline{\mathfrak{Y}})& = [\omega(\underline{\mathfrak{X}}), \omega(\underline{\mathfrak{Y}})]_L + \omega([\underline{\mathfrak{X}}, \underline{\mathfrak{Y}}]_A)\,.
\end{align}
These above defined curvatures are related to each other in the following way \cite{Ciambelli:2021ujl}:
\begin{equation}
\label{Rcurvatures}
R^\sigma(\uX,\uY)=j(\Omega(\umX,\umY))=-j(R^\omega(\umX^H,\umY^H))\,,
\end{equation}
which we have seen in \eqref{curvatureeq} in the \v Cech language. Thus, these notions of curvature actually represent the same thing, namely the curvature of the Lie algebroid.

\subsection{Courant Algebroids}
\label{App:Courant}
A Courant algebroid is a vector bundle $E$ over $M$ equipped with three structures: \ding{172} a bundle map $\rho:E\to TM$ called the anchor map; \ding{173} a symmetric, non-degenerate, fiber-wise bilinear form $\langle\cdot,\cdot\rangle:E\otimes E\to \mathbb{R}$, called the \emph{inner product} or \emph{pairing}; \ding{174} a bilinear operation $[\cdot,\cdot]_{E}:\Gamma(E)\times\Gamma(E)\to\Gamma(E)$ called the \emph{Courant bracket}, which is antisymmetric:
\begin{align}
[\un e_1,\un e_2]_{E} = -[\un e_2,\un e_1]_{E}\,.
\end{align}
These are required to satisfy, for all $\un e_1,\un e_2,\un e_3\in\Gamma(E)$:
\begin{enumerate}[(a)]
\item $\rho([\un e_1,\un e_2]_{E}) = [\rho(\un e_1),\rho(\un e_2)]_{TM}$ (anchor preserves brackets),
\item $[f \un e_1,g \un e_2]_{E} = fg[\un e_1,\un e_2]_{E} + f(\rho(\un e_1)g)\un e_2-g(\rho(\un e_2)f)\un e_1+ \frac{1}{2}\langle \un e_1,\un e_2\rangle (gDf-fDg)$, $\forall f\in C^\infty(M)$, where $D:C^\infty(M)\to\Gamma(E)$ is defined by $\langle Df,e\rangle = \rho(e)(f)$,
\item $\rho(\un e_1)\langle \un e_2,\un e_3\rangle = \langle[\un e_1,\un e_2]_{E},\un e_3\rangle + \langle \un e_2,[\un e_1,\un e_3]_{E}\rangle$,
\item $[\un e_1,[\un e_2,\un e_3]_{E}]_{E} + \text{cyclic} = DT(\un e_1,\un e_2,\un e_3)$, where 
\begin{align}
T(\un e_1,\un e_2,\un e_3)=\frac{1}{3}(\langle[\un e_1,\un e_2]_{E},\un e_3\rangle + \text{cyclic})
\end{align}
is called the Jacobiator.
\end{enumerate}
Condition (d) shows that the Courant bracket does not satisfy the Jacobi identity. This failure motivates the introduction of the Dorfman bracket (also called the Dorfman derivative), which is a bilinear operation $[\cdot\circ\cdot]_{E}:\Gamma(E)\times\Gamma(E)\to\Gamma(E)$ defined by
\begin{align}
[\un e_1\circ \un e_2]_E := [\un e_1,\un e_2]_{E} + \frac{1}{2}D\langle \un e_1,\un e_2\rangle.
\end{align}
The Dorfman bracket satisfies a Leibniz identity (making $(\Gamma(E),[\cdot\circ\cdot]_E)$ a Leibniz algebra):
\begin{align}
[\un e_1\circ[\un e_2\circ \un e_3]_E]_E = [[\un e_1\circ \un e_2]_E\circ \un e_3]_E + [\un e_2\circ[\un e_1\circ \un e_3]_E]_E\,.
\end{align}
In this sense, the Dorfman bracket restores a Jacobi-like identity at the cost of ruining the antisymmetry of the bracket, and the Courant bracket is recovered by antisymmetrization
\begin{align}
[\un e_1,\un e_2]_{E} = \frac{1}{2}([\un e_1\circ \un e_2]_E-[\un e_2\circ \un e_1]_E)\,.
\end{align}
Moreover, the Dorfman bracket is compatible with the anchor and the inner product in a simpler way:
\begin{enumerate}[(a)]
\item $\rho([\un e_1\circ \un e_2]_E) = [\rho(\un e_1),\rho(\un e_2)]_{TM}$,
\item $[\un e_1\circ f \un e_2]_E = f[\un e_1\circ \un e_2]_E + (\rho(\un e_1)f)\un e_2$,
\item $[e\circ e]_E = D\langle e,e\rangle$,
\item $\rho(\un e_1)\langle \un e_2,\un e_3\rangle = \langle[\un e_1\circ \un e_2]\circ \un e_3\rangle + \langle \un e_2\circ[\un e_1\circ \un e_3]\rangle$.
\end{enumerate}
The Courant algebroid can be defined alternatively by replacing the Courant bracket and the corresponding conditions with the Dorfman bracket and the above conditions.

A Courant algebroid is said to be \emph{exact} if the anchor map is surjective and the following exact sequence is satisfied:
\begin{equation}
\label{CourantAppendix}
\begin{tikzcd}
0
\arrow{r}
&
T^*M
\arrow{r}{\rho_E^*}
&
E
\arrow{r}{\rho_E}
&
TM
\arrow{r}
&
0\,,
\end{tikzcd}
\end{equation}
where $\rho^*$ is the pullback of $\rho$ defined through the inner product. In this case, the Courant algebroid is locally identified as $TM\oplus T^*M$. Note that unlike transitive Lie algebroids, the surjectivity of $\rho_E$ does not automatically imply the existence of the short exact sequence \eqref{CourantAppendix}.

\paragraph{Connection and curvature on $E$.} Suppose $E$ is an exact Courant algebroid. As in the transitive Lie algebroid case, there is a canonical vertical subbundle
\begin{equation}
V:=\ker(\rho_E)=\mathrm{im}(\rho_E^*)\subset E,
\end{equation}
which is isotropic with respect to the inner product in the sense that $\langle v_1,v_2\rangle=0$, $\forall v_1,v_2\in V$.

A choice of \emph{connection} (or \emph{split}) is a bundle map $\sigma_E:TM\to E$ satisfying $\rho_E\circ\sigma_E=\mathrm{id}_{TM}$. In general $\sigma_E$ is not unique, and each choice of $\sigma_E$ defines the horizontal subbundle $H=\mathrm{im}(\sigma)$ of $E$, and the Courant algebroid can be decomposed as $E= H\oplus V$. We can also impose the isotropic condition on $\sigma_E$ by requiring that 
\begin{equation}\label{CourantIsotropicSplit}
\langle \sigma_E(\uX),\sigma_E(\uY)\rangle=0\,,\qquad \forall \uX,\uY\in TM\,.
\end{equation}
Then, the horizontal subbundle is said to be \emph{maximally isotropic} or \emph{Lagrangian} (maximal in the sense that it is an isotropic subbundle of largest possible rank).

The curvature of $\sigma_E$ measures the obstruction for horizontal lifts to close under the Courant bracket, defined as
\begin{equation}
R^{\sigma_E}(\uX,\uY)\equiv[\sigma_E(\uX),\sigma_E(\uY)]_E-\sigma_E([\uX,\uY]_{TM})\,.\qquad\forall \uX,\uY\in TM\,.
\end{equation}
Since $\rho_E$ preserves brackets, we have for all $\uX,\uY\in TM$,
\begin{equation}
\rho_E(R^{\sigma_E}(\uX,\uY))=\rho_E\big([\sigma_E(\uX),\sigma_E(\uY)]_E-\sigma_E([\uX,\uY]_{TM})\big)=0\,,
\end{equation}
and hence $R^{\sigma_E}(\uX,\uY)\in \ker(\rho_E)=\mathrm{im}(\rho_E^*)$.
Using the inner product, one can package this vertical component into a 3-form $H\in\Omega^3(M)$ defined by
\begin{equation}
\label{Hsigma}
H(\uX,\uY,\uZ):=\big\langle[\sigma_E(\uX),\sigma_E(\uY)]_E,\sigma_E(\uZ)\big\rangle\,,
\qquad \forall \uX,\uY,\uZ\in TM\,.
\end{equation}
For an isotropic splitting \eqref{CourantIsotropicSplit}, $H$ is totally antisymmetric and one has the identity
\begin{equation}\label{CourantCurvIdentity}
[\sigma_E(\uX),\sigma_E(\uY)]_E=\sigma_E([\uX,\uY]_{TM})+\rho_E^*(i_{\uY}i_{\uX}H)\,,
\end{equation}
which makes explicit that $H$ is the curvature of the exact Courant algebroid. Moreover, $H\in \Omega^3(M)$ is a closed 3-form. The cohomology class $[H]\in H^3(M,\mathbb{R})$ is the Ševera class, classifying exact Courant algebroids up to isomorphism. 

\section{Batalin--Vilkovisky (BV) Formalism and Quantization of Abelian 2-Form Gauge Fields}
\label{App:BV}
Following~\cite{Henneaux:1992ig}, we present the BV quantization of a free theory of the Abelian two-form gauge field $B_{\mu\nu}$ in spacetime dimensions $d \geqslant 6$. For $d=4$ and $d=5$ the free two-form field is dual to compact scalar and $U(1)$ gauge field respectively, and it is hence economical to go to the dual frame and perform the quantization there in the conventional manner. For spacetime dimensions $d \leqslant 3$, the free two-form field cannot carry any local physical degrees of freedom.

On a local patch, the free theory of the Abelian 2-form $B_{\mu\nu}$ is described by the classical action 
\begin{equation} \label{eq:S0-twoform}
    S_0[B] = -\frac{1}{12} \int \td^dx\, H_{\mu\nu\rho} H^{\mu\nu\rho}\, ,
\end{equation}
expressed in terms of its field strength defined by $H_{\mu\nu\rho} = 3 \partial_{[\mu} B_{\nu\rho]}$.  $H$ and the action are invariant under the \emph{reducible}~\cite{Batalin:1984jr} gauge transformation 
\begin{equation} \label{eq:gaugetranformationB}
\delta B_{\mu\nu} =  2 \partial_{[\mu} \Lambda_{\nu]} \,, 
\end{equation}
where there are \emph{gauge-for-gauge} transformations $\delta \Lambda_\mu = \partial_\mu \lambda$. This indicates that the gauge transformations \eqref{eq:gaugetranformationB} are not linearly independent.
The gauge parameter $\Lambda_\mu$ has $n = d$ components and $\lambda$ has $m=1$ components, for a total of $n-m=d-1 $ independent gauge symmetries. We will amply see that the gauge fixing procedure would respect exactly this number.

Since gauge-for-gauge transformations terminate at the first round, this type of gauge symmetry is called \emph{first-stage reducible} according to the Batalin--Vilkovisky (BV) field-antifield formalism \cite{Batalin:1981jr, Batalin:1984jr}.  Whenever reducible gauge symmetries appear, the traditional BRST quantization is not sufficient and one needs to adopt the BV quantization. In the following we demonstrate how to apply the BV algorithm to $B_{\mu\nu}$.

\paragraph{Extended field space.} 
We introduce \emph{ghost fields} $\{C_\mu, \, c \}$ for gauge parameters $\Lambda_\mu$ and $\lambda$.
\emph{Grassmann parity} for $B_{\mu\nu}, \Lambda_\mu$ and $\lambda$ are $\epsilon(B_{\mu\nu}) = \epsilon(\Lambda_\mu) = \epsilon(\lambda)= 0\mod 2$ and as for the ghosts, they are always parity-opposed to their gauge parameters: $\epsilon(C) = \epsilon(\Lambda) + 1$ and $\epsilon(c) = \epsilon(\lambda) + 1$. We have ghosts from two different stages. To distinguish between them, we introduce the \emph{ghost number} denoted as $\gh(X)$ for a field $X$. For our classical fields $X\in\{B_{\mu\nu}, \Lambda_\mu,\lambda\}$, we have $\gh(X) = 0$ whereas for the ghosts they are just the stage number plus one, i.e., $\gh(C) = 1$ and $\gh(c) = 2$.

We then need to double the space of field variables by assigning to each field $\Phi^I$ (collective notation for the set of all fields) its \emph{antifields} $\Phi^*_I$.
The ghost number assignments and Grassmann parities of the antifields are related to their ordinary counterparts
\begin{equation}
   \gh(\Phi^*_I) = - \gh(\Phi^I) - 1 \, ,\quad \epsilon(\Phi^*_I) = \epsilon(\Phi^I) + 1 \quad \mod 2\, .
\end{equation}
In addition, Grassmann parity also equals the ghost number modulo two.
In such a way, we arrive at the \emph{minimal} sector
    \begin{equation}
    \{ B_{\mu\nu}\, , C_\mu\, , c\, ; B^{*\mu\nu}\, , C^{*\mu}\, , c^* \}\, .
\end{equation}
The action $S_0[B]$ is then augmented to the minimal BV action $S_{\text{Min}}[\Phi,\Phi^*]$
depending on the original fields but also on the ghost fields and their antifields. It is a ghost number zero, even functional that should be a proper solution of the \emph{classical master equation}
\begin{equation}
(S_{\text{Min}}, S_{\text{Min}}) = 0\, ,
\end{equation}
where the \emph{antibracket} $(\, \cdot \, , \, \cdot \,)$ on the extended field spaces is defined as\footnote{We use the superscripts ``$L$'' and ``$R$'' to indicate whether the functional derivative is acting from the left or from the right, respectively. For any function or functional $X$ of the field $\phi$ we have the variation $\delta X(\phi)=\delta \phi \frac{\delta^{L} X}{\delta \phi}=\frac{\delta^{R} X}{\delta \phi} \delta \phi$.}
\begin{equation}
    (X,Y) = \frac{\delta^R X}{\delta \Phi^I}\frac{\delta^L Y}{\delta \Phi^*_I} - \frac{\delta^R X}{\delta \Phi^*_I}\frac{\delta^L Y}{\delta \Phi^I}\, .
\end{equation}
Moreover, it should satisfy the boundary condition:
\begin{equation}
S_0[B] = S_{\text{Min}}[\Phi, \Phi^* = 0]\, .
\end{equation}
In our case, minimal action reads
\begin{equation}
    S_{\text{Min}} = \int \td^dx\, \left( -\frac{1}{12} H_{\mu\nu\rho} H^{\mu\nu\rho} + 2 B^{*\mu\nu} \partial_{\mu} C_\nu + C^{*\mu} \partial_\mu c \right)\,.
\end{equation}
To gauge-fix the gauge freedom of $C_\mu$ and $\bar{C}^\mu$ themselves, one adds the usual antighost part, $\{ \bar{C}^\mu\, , b^\mu \}$ (+ their antifields), but also two more pairs $\{ \bar{c}, b \}$ and $\{ \eta, \pi \}$ (+ their antifields).  The minimal sector, together with these additional fields and antifields, is called the \emph{non-minimal} sector whose action also satisfies the master equation and it is just 
\begin{equation} \label{eq:SNM-2form}
    S_{\text{NM}} = S_{\text{Min}} + \int \td^dx \left( \bar{C}^*_\mu b^\mu + \bar{c}^* b + \eta^* \pi \right)\,.
\end{equation}
Ghost numbers are as follows:
\begin{equation}
    \begin{tabular}{c|cccccc|ccc}
  & $B_{\mu\nu}$ & $C_{\mu}$ & $\bar{C}^\mu$ & $c$ & $\bar{c}$ & $\eta$ & $b^\mu$ & $b$ & $\pi$ \\ \midrule
gh & 0 & 1 & -1 & 2 & -2 & 0 & 0 & -1 & 1
\end{tabular}\, .
\end{equation}
  With these rules, the BV \emph{master actions} (minimal and non-minimal) are even and of ghost number zero.
\paragraph{BRST--BV gauge-fixing.} An appropriate \emph{gauge-fixing fermion} is
\begin{equation} \label{delta-gauge-fixing-1}
    \Psi_{\delta} = \int\td^dx \left( \bar{C}^\mu \partial^\nu B_{\nu\mu} + \bar{c}\, \partial^\mu C_\mu +\bar{C}^\mu \partial_\mu \eta \right)\, .
\end{equation}
It is odd and of ghost number $-1$. It gives the gauge-fixed action 
\begin{align}
    S_\delta [\Phi^I] &= S_{\text{NM}}\left[\Phi^I, \Phi^*_I =  \frac{\delta \Psi}{\delta \Phi^I}\right] \\
    &= \int\td^dx \Big( -\frac{1}{12} H_{\mu\nu\rho} H^{\mu\nu\rho} -\frac{1}{2} \bar{F}^{\mu\nu} F_{\mu\nu} - \partial_\mu \bar{c}\, \partial^\mu c \nonumber\\
    &\qquad\qquad + \left( \partial^\nu B_{\nu\mu} + \partial_\mu \eta \right) b^\mu + (\partial^\mu C_\mu) b - (\partial_\mu \bar{C}^\mu) \pi \Big)\, ,
\end{align}
which is independent of the antifields. We used the notation
\begin{equation}
    \bar{F}^{\mu\nu} = 2 \partial^{[\mu} \bar{C}^{\nu]}\, , \quad F_{\mu\nu} = 2 \partial_{[\mu} C_{\nu]}\, ,
\end{equation}
which makes it clear that the ghosts themselves have the gauge symmetry $\delta C_\mu = \partial_\mu \alpha$, $\delta \bar{C}^\mu = \partial^\mu \bar{\alpha}$ (in the absence of the second line of the action of course). Up to a factor, they have the normal kinetic term for a complex vector field (with the wrong Grassmann parity, as expected).

The equations of motion for the auxiliary fields give
\begin{align}
    \partial^\nu B_{\nu\mu} + \partial_\mu \eta = 0 \, , \quad \partial^\mu C_\mu = 0\, , \quad
    \partial_\mu \bar{C}^\mu = 0\, .
\end{align}
For the ghost fields, this is the usual Lorenz gauge condition. For the two-form, the left-hand side is not identically divergenceless anymore (which would give a $\delta(0)$ in the path integral). Still, it implies $\Box \eta = 0$, which with appropriate boundary conditions gives $\eta = 0$ and so $\partial^\nu B_{\nu\mu}=0$.
We find also that $\partial_{\mu} ( \partial^\nu B_{\nu\mu}) =0$ identically and it hence gives the correct number $n-m = d-1$ of gauge conditions to fix the independent gauge transformations.

\paragraph{$\delta$-function vs Gaussian gauge-fixing}
The gauge system is first-stage reducible, and we performed the $\delta$-function gauge fixing \eqref{delta-gauge-fixing-1} procedure to obtain a gauge fixed action $S_\delta [\Phi^I]$. In fact, one can have multiple consistent choices of gauge fixing fermions. Here we choose another $\Psi$ by adding to the $\delta$-function gauge fixing fermion the following functional
\begin{equation} \label{Gaussian-gauge-fixing-1}
    \Psi_{\pi} = \int\td^dx \left( \frac{1}{2\xi} \bar{C}^\mu b_{\mu} + \frac{1}{2\kappa} \bar{c}\, \pi - \frac{1}{2\kappa} \eta b \right)\, ,
\end{equation}
which is linear in the auxiliary fields $b_{\mu}$, $b$ and $\pi$. We also put two parameters $\xi$ and $\kappa$ in the gauge-fixing fermion. We then eliminate the antifields with the new gauge-fixing fermion
 \begin{equation}
 \Psi = \Psi_{\delta} + \Psi_{\pi}.     
 \end{equation}
The gauge-fixed action is
\begin{align} \label{gaussian-gauge-fixing-1}
    S_{\text{g.f.}}[\Phi^I] &= \int\td^dx \Bigg( -\frac{1}{12} H_{\mu\nu\rho} H^{\mu\nu\rho} -\frac{1}{2} \bar{F}^{\mu\nu} F_{\mu\nu} - \partial_\mu \bar{c}\, \partial^\mu c \nonumber\\
    &\qquad\qquad + \left( \partial^\nu B_{\nu\mu} + \partial_\mu \eta + \frac{1}{2\xi} b_{\mu} \right) b^\mu + \left(\partial^\mu C_\mu  + \frac{1}{2\kappa} \pi\right) b - \left(\partial_\mu \bar{C}^\mu + \frac{1}{2\kappa} b \right) \pi \Bigg)\, ,
\end{align}
which is ready for performing the path integral.
One can first path integrate over the auxiliary fields $b_{\mu}$, $b$ and $\pi$. After integration, there are new terms in the action that contribute to the propagators. This part of the action is also called gauge-breaking contribution \cite{Batalin:1984jr}
\begin{equation}
     S_{\text{g.b.}}[\Phi^I] = \int\td^dx \Big( -\frac{\xi}{2} ( \partial^{\alpha} B_{\alpha \mu})( \partial_{\beta} B^{\beta \mu} )
     + \kappa\, \partial_{\mu} \bar{C}^{\mu} \partial_\nu C^\nu - \frac{\xi}{2} \partial_{\mu} \eta \partial^{\mu} \eta \Big)\, ,
\end{equation}
The first term in the gauge-breaking part ensures that the bosonic two-form $B$ has a non-degenerate propagator. The second term also removes the degeneracy of the $\bar{F}F$ term. The last term indicates that the ghost for ghost field $\eta$ is propagating. 
No cross-term between $B$ and $\eta$ appears here, since $(\partial^\nu B_{\mu\nu}) \partial^\mu \eta$ is a total derivative. This is equivalent to the constraint $\partial^\mu F_\mu = 0$ on the gauge-fixing function $F_\mu \equiv \partial^\nu B_{\mu\nu}$.

\paragraph{Counting of degrees of freedom.}

We must count the non-auxiliary fields, with alternating signs.
The 2-form accounts for
\begin{equation}
    N_2 = \frac{d(d-1)}{2}\,,
\end{equation}
the pair $(C_\mu, \bar{C}^\mu)$ gives
\begin{equation}
    N_1 = 2 d\,,
\end{equation}
and $(c, \bar{c}, \eta)$ gives
\begin{equation}
    N_0 = 3\,,
\end{equation}
where the subscript indicates the number of indices. In total, this gives
\begin{equation}
    N = N_2 - N_1 + N_0 = \frac{(d-2)(d-3)}{2}
\end{equation}
physical degrees of freedom as it should.

\paragraph{BRST transformations.}
The gauge-fixed actions above are invariant under a nilpotent BRST transformation of ghost number $+1$, and the extra terms in the action (gauge-breaking terms and ghosts terms) are BRST-exact. This comes very naturally out of the field-antifield formalism; we refer to the reviews \cite{Henneaux:1992ig,Gomis:1994he} for a general discussion.

In our case, the action of the BRST differential $\ts$ on a functional $A$ depending on the fields $\Phi^I$ of the non-minimal sector (but not on the antifields $\Phi^*_I$) is given by
\begin{equation}
    \ts A = (A, S_{\text{NM}})\Big\vert_{ \Phi^* = \frac{\delta \Psi}{\delta \Phi} }\, = \left.\frac{\delta^R A}{\delta \Phi^I}\frac{\delta^L S_{\text{NM}}}{\delta \Phi^*_I}\right|_{\Phi^* = \frac{\delta \Psi}{\delta \Phi}}\, .
\end{equation}
Notice, however, that \eqref{eq:SNM-2form} is linear in the antifields\footnote{Terms of higher order in antifields would be expected in a putative interacting theory with a more involved gauge structure, e.g.~if the gauge algebra were open.}; therefore, $\frac{\delta^L S_{\text{NM}}}{\delta \Phi^*_I}$ is antifield-independent and the definition of $s$ in fact does not depend on the gauge-fixing fermion. On the fields, $s$ explicitly reads
\begin{align}\label{eq:BRST}
    \ts B_{\mu\nu} &= 2 \partial_{[\mu} C_{\nu]}\, , \quad \ts C_\mu = \partial_\mu c\, , \quad \ts \bar{C}_\mu = b_\mu\, , \quad \ts \bar{c} = b \, , \quad \ts \eta = \pi \, , \quad \ts (\text{other}) = 0 \, .
\end{align}
The nilpotency 
\begin{equation}
\ts^2 = 0
\end{equation}
is immediate and holds off-shell. On $B_{\mu\nu}$ (and $C_\mu$ due to the reducibility), $\ts$ takes of course the familiar form ``gauge transformations with parameter replaced by ghost''. The gauge-fixed action \eqref{gaussian-gauge-fixing-1} can then be written as
\begin{equation}
    S_{\text{g.f.}} = S_0 + \ts \Psi\, ,
\end{equation}
with $S_0$ the original action \eqref{eq:S0-twoform}. This can be checked explicitly using formulas \eqref{eq:BRST}, or proven more abstractly as follows: since $S_{\text{NM}}$ is linear in antifields, we have $S_{\text{NM}} = S_0 + \Phi^*_I \frac{\delta^L S_{\text{NM}}}{\delta \Phi^*_I}$. Therefore,
\begin{align}
    S_{\text{g.f.}} = S_{\text{NM}} \left[ \Phi^I, \Phi^*_I = \frac{\delta \Psi}{\delta \Phi^I} \right] = S_0 + \frac{\delta \Psi}{\delta \Phi^I} \frac{\delta^L S_{\text{NM}}}{\delta \Phi^*_I} = S_0 + \ts \Psi\, .
\end{align}
BRST invariance
\begin{equation}
    \ts S_{\text{g.f.}} = 0
\end{equation}
of the gauge-fixed action then follows from the gauge-invariance of $S_0$ (indeed, $\ts S_0 = 0$ is equivalent to its gauge invariance since it only depends on $B_{\mu\nu}$) and $\ts^2 = 0$.

\paragraph{Russian formula and the BV--BRST complex.}
A convenient way to encode at once the gauge symmetry \eqref{eq:gaugetranformationB}, its first-stage reducibility, and the resulting BRST algebra is through the \emph{Russian formula}. Conceptually, this identity is already the core input of the BRST description of ordinary (irreducible) gauge theories: it states that one can extend the de~Rham differential $\td$ by the BRST differential $\ts$ so that the curvature is totally horizontal, i.e.~it has no ghost components.  For a first-stage reducible system such as the Abelian two-form, the same idea persists but requires including the ghost-for-ghost; we may refer to the resulting identity as the \emph{higher Russian formula}.

In the free Abelian case, the differentials $\td$ and $\ts$ define a bicomplex
\begin{equation}
\td^2=0\,,\qquad \ts^2=0\,,\qquad \td\ts+\ts\td=0\,,
\end{equation}
called the BV--BRST complex. Introduce the extended field
\begin{equation} 
\hat{B} \equiv B - C + c \, ,
\end{equation}
with $\gh(B,C,c)=(0,1,2)$.  The (higher) Russian formula is the horizontality requirement
\begin{equation}\label{eq:russian-formula}
(\td+\ts)\,(B - C + c) = H \equiv \td B \, ,
\end{equation}
namely the extended differential $\td+\ts$ has no ghost contributions in the curvature.
Expanding \eqref{eq:russian-formula} by form degree (or equivalently by ghost number) gives
\begin{equation}
\ts B - \td C + (\td c - \ts C) + \ts c = 0\, ,
\end{equation}
and since the three bracketed terms have different form degrees, they must vanish separately:
\begin{equation}\label{eq:brst-from-russian}
\ts B = \td C\,,\qquad \ts C = \td c\,,\qquad \ts c = 0\, .
\end{equation}
Thus, imposing the higher Russian formula \eqref{eq:russian-formula} automatically generates the BRST transformations, while the BV master action constructed above provides their canonical realization via antifields.

\providecommand{\href}[2]{#2}\begingroup\raggedright\endgroup

\end{document}